\documentclass[aps,nofootinbib,preprint,tightenlines]{revtex4}

\usepackage{graphicx}
\usepackage{amsmath}
\usepackage{bm} 
\usepackage{physics}
\usepackage{mathrsfs}
\usepackage{xcolor}

\newcommand{\beq}{\begin{eqnarray}}
\newcommand{\eeq}{\end{eqnarray}}
\newcommand{\bqa}{\begin{eqnarray}}
\newcommand{\eqa}{\end{eqnarray}}

\begin{document}


%
\title{%
A short-range effective theory for single-neutron halo nuclei with a deformed core
}
\author{L.-P. Kubushishi}\email{lkubushi@ohio.edu}
\affiliation{Department of Physics \& Astronomy and Institute of Nuclear \& Particle Physics, Ohio University, Athens, Ohio 45701, USA}
\author{D.~R.~Phillips}\email{phillid1@ohio.edu}
\affiliation{Department of Physics \& Astronomy and Institute of Nuclear \& Particle Physics, Ohio University, Athens, Ohio 45701, USA}

\begin{abstract}
We establish a short-range effective theory for deformed s-wave halos. The theory applies to a system in which neutrons are weakly bound to a core nucleus, and that core nucleus also exhibits a low-lying rotational band with a $0^+$ ground state and a first $2_1^+$ excited state. The effective theory then must have both halo degrees of freedom and degrees of freedom associated with rotation of the core. This leads to the particle plus rotor model of Bohr and Mottelson at leading order. We  identify the relevant leading-order operators in the Hamiltonian that respect the symmetries of the system and the small parameters which define the effective theory's power counting. We carry out calculations for the  $^{11}$Be and $^{17}$C systems in which we compute the low-lying positive-parity states of the core $+$ neutron systems up to the core$(2_1^+)$-neutron threshold. We do this for several different regulator parameters and establish that these energies can be renormalized using the leading-order set of operators. The spectrum is accurately described at leading order in both cases. Decay widths to d-wave core-neutron states have sizable cutoff dependence, but the Asymptotic Normalization Coefficients (ANCs) in $s$-wave channels exhibit regulator dependence of a size consistent with next-to-leading-order effects. We also compute Coulomb breakup observables and compare with experimental data, finding leading-order results in reasonable agreement with data for both ${}^{11}$Be and ${}^{17}$C. The addition of one next-to-leading-order operator renders the ANCs of all s-wave states stable with respect to the regulator. It consequently also removes most of the  30\% variation  of the leading-order Coulomb dissociation cross section with the regulator.
\end{abstract} 
\maketitle
\section{Introduction}
The existence of halo nuclei is a striking manifestation of quantum mechanics. Their characteristic large size is due to loosely bound valence nucleons tunneling far outside the mean-field potential generated by the halo's compact core. Since their discovery in the 1980s \cite{Tan85b,Tan85l}, halo nuclei, which exhibit a significantly larger size than their neighbors in the nuclear chart, have  attracted a lot of attention, both experimentally and theoretically \cite{HJ87,Tan96}.

Early studies of these systems helped define the notion of a halo, the relation between weak binding and quantum tunneling, and point out halos' clusterized structure \cite{HJ87,RIIS94,Tan96}. Once the fact that halos could be treated as effective few-body---core + neutron(s)---problems was highlighted, a wide variety of effective core-neutron potential models were developed to better understand them and, in particular, predict and analyze data from reactions involving halo nuclei \cite{THOMPSON88,MISU97,JEN04,KF16,KF162,JIM17,Moro25,MARIDI25}. More recent theoretical approaches incorporated many-body correlations and continuum effects to explore heavier halos, including $^{29}$Ne, $^{31}$Ne, $^{37}$Mg, and $^{42}$Al~\cite{LI22,ZHANG23prc,ZHANG23plb}. Other studies of binding energy systematics suggested that one- and two-nucleon halos near the neutron and proton drip lines, may be emitted in the decay of superheavy nuclei~\cite{SANT2018,ANJA2020,ANJA2020b}. Experimentally, efforts have continued to extend the search for new and heavier halo systems, such as $^{31}$Ne, $^{37}$Mg, and $^{29}$F~\cite{NAKA14,KOBA14,TAKE14,BAGCHI2020}. Other experiments have also introduced new observables to probe halo structure, including the ratio method~\cite{OTA2025} and measurements of dineutron and diproton correlations in two-nucleon halo systems~\cite{LEHR2022,CORSI2023}. 

Effective field theories (EFTs) have provided a systematic and model-independent approach to low-energy nuclear phenomena \cite{HAM17}. In particular, EFTs with short-range interactions make explicit the fine tuning that is needed to produce a weakly bound system, and have been applied to shallow bound states in the nuclear, particle, and atomic realms~\cite{KSW98A,KSW98B,VANKOLCK99,BRM99,BH04}. Halo nuclei are a natural target for such an approach, which takes advantage of their clear separation of scales. Halo EFT  embeds the clusterized structure of halo nuclei in a controlled expansion in powers of $R_{\rm core}/R_{\rm halo}$, where $R_{\rm core}$ is the size of the core nucleus, and $R_{\rm halo}$ is the large distance scale associated with the halo nucleon(s)~\cite{BERT02,BED03}. This has been typically done under the assumption of an inert core and  has been particularly successful in analyzing and describing reaction observables of halo nuclei \cite{CAP18,YC18,MYC19,HC21}. EFTs for certain halos that include a low-lying excited state of the core nucleus as a dynamical degree of freedom (``core excitation'')  have also been proposed~\cite{RUP11,ZNP15}.

Low-lying excited states of the core are to be expected in deformed nuclei. There too there is a separation of scales, especially as the mass of the nucleus, and hence its moment of inertia, increases. In such a nucleus the energy required to excite a collective rotational mode of the entire nucleus can be much smaller than that associated with single-particle excitations. EFTs that exploit this separation of scales have been developed for both even-even and odd-even nuclei~\cite{PAP16,PAP20,ALCPP21,ALCPP22}. The expansion here is in $E_{\rm rotor}/E_{\rm hi}$, where $E_{\rm hi}$ is the energy scale of the first degree of freedom to appear beyond rotations: vibrations, pair breaking, single-particle excitations, or something else.  

But what of systems which are both deformed {\it and} have a halo nucleon? In this manuscript we establish an effective theory for such nuclei: a short-range EFT for deformed halo nuclei. 
We build the leading order of this theory for s-wave neutron halos, whose core is deformed. Such cores exhibit a low-lying rotational band  with a $0^+$ ground state and a first $2_1^+$ excited state. We start with Halo EFT and add the low-energy rotational degree of freedom of the core to the theory. We establish the relevant operators that respect the symmetries of the system and contribute to the Hamiltonian at leading order. We also establish the power counting, i.e., the amount by which non-leading operators containing  halo and/or rotational degrees of freedom are suppressed. We use the EFT to compute the low-lying positive-parity states of
core + neutron systems up to the core($2^+_1$) threshold for several different regulator parameters. We discuss the renormalization of observables, such as the binding energies and ANCs, calculate Coulomb breakup observables, and compare with experimental data. All of this done for the one-neutron halos $^{11}$Be and $^{17}$C.

This manuscript is organized as follows. In Section~\ref{sec:deformedSRET}, we review the basics of Halo EFT and then describe how rotational degrees of freedom can be incorporated in it and, how, at leading order, this leads to the particle plus rotor model of Bohr and Mottelson \cite{BOMOT69}. We identify the relevant leading-order operators contributing to the Hamiltonian of the system, define the power counting, and discuss the different regimes that are possible depending on the hierarchy of rotational and halo scales. In Sec.~\ref{impl_LO} we describe the ingredients and method we use to solve for EFT energy levels as well as computing Coulomb breakup observables. For $^{11}$Be and $^{17}$C, the obtained results, such as spectra up to the core($2^+_1$) threshold and Coulomb breakup observables, are then presented and analyzed in Sec.~\ref{sec:cases} for different regulator parameters. In Sec.~\ref{sec:beyondlo} we study the impact of one beyond-leading operator on Coulomb observables and discuss renormalization. Finally, we conclude with a summary and discussion of future pathways for work in this EFT.
\section{Short-range effective theory with deformed halos}\label{sec:deformedSRET}

\subsection{Halo EFT with a spherical core at leading order}\label{sec:haloeft}

The interaction between neutrons and a spherical core is described by an extension of the pionless EFT~\cite{BRM99,CHEN99,VANKOLCK99} called ``Halo EFT''~\cite{BERT02,BED03}, see the review~\cite{HAM17}.  In this EFT the Lagrangian is an expansion in powers of the small parameter
\begin{equation}\label{expPARAM1}
Q \equiv \frac{R_{\rm core}}{R_{\rm halo}}, k R_{\rm core} \equiv\frac{M_{\rm halo}}{M_{\rm core}},  \frac{k}{M_{\rm core}};
\end{equation}
where $R_{\rm X}$ ($M_{\rm X}$) are the distance (mass) scales associated with the core and halo of the nucleus and we adopt the standard estimate $R_{\rm X} \sim 1/M_{\rm X}$. This EFT is therefore designed for systems that are dominated by the asymptotic part of their wave function, and so have an rms radius $R_{\rm halo}$ that is much larger than the range of the interaction $R_{\rm core}$. It will fail once the system is probed at wave numbers $k$ such that $k R_{\rm core} \sim 1$. We note that in this picture the substructure of the core is not resolved and enters the Lagrangian only through higher-dimensional operators of $O(Q^2)$ and beyond. 

The core and neutron's free motion are described by the usual non-relativistic kinetic operator. This operator is tuned against the potential-energy operator, which is irrelevant according to naive dimensional analysis. The amount of fine tuning required to form a weakly bound, i.e. halo, state is minimized for a system of particles containing at most one charged object and for $s$-wave interactions. At leading order (LO) in Halo EFT shallow s-wave bound states are described by a contact core-neutron potential
\begin{equation}
   \rm{V^{LO}_{cn}=C_0\delta^{(3)}(r)},
    \label{eq1}
\end{equation}
where $\rm{C_0}$ is a low-energy constant which is adjusted to reproduce  the binding energy of the real or virtual bound state in the neutron-core halo, or, equivalently (up to this order in the expansion) the neutron-core scattering length. 
The operator (\ref{eq1}) is an irrelevant operator according to naive dimensional analysis. However, if this potential is to 
produce a shallow bound state the coupling $\rm{C_0}$ must appear at leading order and must be fine tuned against the kinetic energy operator. This results in an expansion for operators associated with the halo degree of freedom that proceeds in powers of $Q$, not in powers of $Q^2$~\cite{KSW98A,KSW98B,BRM99,VANKOLCK99}.

The single-particle core-neutron Hamiltonian at LO in Halo EFT is then
\begin{equation}
   \rm{\hat{H}^{LO}_{halo-EFT}(\textbf{r})=\hat{T}_{\textbf{r}}+V^{LO}_{cn}(r)},
    \label{eq:hamiltHaloEFTlo}
\end{equation}
with $\rm{\hat{T}_{\textbf{r}}}$ the single-neutron kinetic energy operator and ${\bf r}$ the  core-neutron relative co-ordinate. 

\subsection{Halos with a deformed core}

Now, suppose that, instead of the spherically symmetric system implicit in the discussion of the previous subsection, we are dealing with a deformed core.
We assume, following Bohr and Mottelson \cite{BOMOT69}, that, in the frame in which it is at rest, the core undergoes a permanent quadrupole deformation and so has axial symmetry. In this ``intrinsic'' frame we can then choose the $z$-direction to be the one that breaks spherical symmetry and parameterize the breaking of the symmetry by an elongation factor $\zeta$.  
 The EFT for spherical halos can therefore be mapped to an EFT for axially symmetric halos by applying to the EFT interaction the operation
\begin{equation}
r^2 \rightarrow \frac{x^2 + y^2}{R^2_{\perp}} + \frac{z^2}{R^2_{z}},
\label{eq:deform}
\end{equation}
where $R_{z}=\zeta R_0$, $R_\perp$ is the extent of the system in the $x-y$ plane, and $R_0 \sim R_{\rm core}$ denotes the radius of the undeformed spherical core. Hence, $\zeta > 1$ describes an elongation along $z$ (prolate core) and $\zeta < 1$ a shrinking along $z$ (oblate core). The modification of the radius along the transverse plane $R_{\perp}$ is constrained to ensure volume conservation 
\begin{equation}
    \frac{4}{3}\pi R_0^3=\frac{4}{3}\pi R^2_{\perp}R_z \iff R_{\perp}=\frac{R_0}{\sqrt{\zeta}}.
    \label{volconserv}
\end{equation}

By introducing the parameter
\begin{equation}
\beta \equiv 2\left(\zeta-1\right),
\end{equation}
we can rewrite the volume-preserving mapping as
\begin{equation}
{\cal D} r^2=\left[1 - \beta P_2(\cos \theta) \right]r^2,
\end{equation}
where $P_2$ is the second Legendre polynomial, and the operator ${\cal D}$ generates the action of the deformation transformation (\ref{eq:deform}) on the object it is acting on. 

Expanding in powers of $\beta$ we have that the action of ${\cal D}$ on a spherically symmetric EFT interaction $V(r^2)$ is
\begin{equation}
{\cal D} V(r^2)=V(r^2) - \beta P_2(\cos \theta) r^2 \frac{d V}{d (r^2)} + O(\beta^2) .
\label{eq:deformedcore}
\end{equation}
When applied to the Halo EFT potential (\ref{eq1}) this generates a leading-order short-distance interaction which---to linear order in $\beta$---accounts for the manner in which the deformed core breaks the spherical symmetry of the core-neutron potential and reduces it to an axial symmetry. 
Crucially, this deformation operator does not alter the short-distance behavior of $V$, i.e., it does not introduce new divergences in any graph in which $V$ acts. 

On the face of it, the action of ${\cal D}$ appears to violate rotational invariance, since it creates a dependence of the interaction on $\cos \theta$. And the interaction is non-central, so the square of the angular momentum associated with the neutron-core relative co-ordinate is no longer a good quantum number. However, $\theta$ here is defined in the intrinsic frame. More generally, it is the angle between the vector ${\bf r}$ and the vector that describes the orientation of the core spin, {\bf I}. Replacing $\cos \theta$ by $\rm{\hat{\bf{I}} \cdot \hat{r}}$ makes it clear that the interaction is still invariant under global rotations. The non-central $c$-$n$ interaction, and this argument, is the basis of the rigid-rotor-plus-particle model \cite{BOMOT69}. 
\subsection{Activating rotational degrees of freedom}\label{sec:rotoreft}
We now assume further that the axially-deformed core can be described at leading order as a rigid rotor. Its $0^+$ ground state is the bandhead of the rotational band, which has a first $2^+_1$ excited state. The EFT description of the deformed-core-neutron interaction must therefore incorporate these low-lying even-parity states associated with the core's collective rotations. 

In this EFT the leading-order intrinsic Hamiltonian of the core is~\cite{PAP16}
\begin{equation}
\rm{H^{LO}_{rotor}}=\frac{\mathbf{I}^2}{2\vartheta},
\label{eq:LOHrotor}
\end{equation}
where ${\bf I}$ is the spin of the core and $\vartheta$ is its moment of inertia relative to any axis other than the axis of symmetry. The EFT of nuclear rotations organizes corrections to the rigid-rotor picture. This organization is clearest if one begins with the Lagrangian, which is written in terms of the rotational velocity $\rm{v}$. Since we consider only the first excited $2_1^+$ state of the core, we have ${\rm I} \sim 1$, and the moment of inertia $\vartheta$ scales as $1/{\rm v}$:
\begin{equation}
    \rm{I=\vartheta \left| \hat{v} \right| \sim 1} \Rightarrow \rm{\vartheta \sim 1/v}.
    \label{eq:frequencyscaling}
\end{equation}
The EFT then expands the rotor Lagrangian in powers of the small parameter
\begin{equation}
\rm{\xi} \equiv \frac{\rm{v}}{\rm{v_{hi}}},
\end{equation}
with $E_{\rm rotor}=\frac{1}{2} \vartheta {\rm v}^2$, and $\rm{v_{hi}}$ the angular velocity that is sufficient to break the rigid rotor, i.e., ${\rm v}_{\rm hi}$ the angular velocity associated with vibrational degrees of freedom, pair breaking, the first shell-model excitation of the core, etc. Such corrections to the leading-order Lagrangian---corrections that represent non-rigidity of the core---are then $O(\xi^2)$. They map to sub-leading corrections $\sim I^4$ to the leading-order rotor Hamiltonian (\ref{eq:LOHrotor})~\cite{PAP16,PAP20}.

The additional degree of freedom in the low-energy Lagrangian associated with rotations, ${\rm v}$, can also interact with the degrees of freedom associated with the halo neutron to generate core-neutron operators, i.e., pieces of the core-neutron potential that depend on ${\rm v}$. The interactions that can be generated were organized in an expansion in powers of $\xi$ by Alnamlah, Coello Perez, and Phillips in Ref.~\cite{ALCPP21}, although that work expressed the Hamiltonian as a sum of one-body operators, organized in powers of the overall angular momentum of the fermion-rotor state. 

Here we aim at a two-body description of rotor-fermion interaction and so we instead use the rotor angular momentum, ${\bf I}$ as the rotational degree of freedom. There is a core-neutron operator of $O({\rm v})$ (and so linear in ${\bf I}$) and is permitted by the symmetries of the problem. It is:
\begin{equation}
\rm{H^{int}_{HF}}=C_1 \bm{\sigma} \cdot \mathbf{I} \delta^{(3)}(r),
\label{eq:HF}
\end{equation}
and it generates a hyperfine coupling between the core and the valence neutron. We take it to be a zero-derivative contact interaction so that we work to leading-order in the Halo EFT expansion of this interaction.  The coefficient $\rm{C_1}$ in Eq.~(\ref{eq:HF}) is undetermined; it is not protected by rotational symmetry. The impact of $\rm{C_1}$ on halo nuclei's sprectra is expected to be of the order of the gap between the low-energy $\frac{3}{2}^+$ and $\frac{5}{2}^+$ states, which is in turn expected to be of the same order as the energy gap between the $0^+$ and $2_1^+$ of the core. Consequently, the hyperfine operator enters at leading order in our EFT. As we shall see below, at leading order, this operator is not needed to renormalize binding energies. We include it for phenomenological reasons as without it, experimental level ordering is not reproduced correctly and $\frac{3}{2}^+$ and $\frac{5}{2}^+$ states are degenerate.

Operators with the structure $\sim  \bm{\sigma} \cdot \mathbf{I}$ that are associated with non-rigidity of the core enter the rotor-fermion EFT Lagrangian at order ${\rm v}^3$, and so are suppressed by $\xi^2$ compared to this LO hyperfine interaction. Meanwhile there are also corrections to (\ref{eq:HF}) that involve derivatives of the halo fermion degree of freedom: these are suppressed by only one power of $Q$ if the coefficient $\rm{C_1}$---like the coefficient $\rm{C_0}$---is fine tuned.

The leading-order Hamiltonian for the core-neutron halo system is therefore
\begin{equation}
\begin{split}
\hat{\rm H}^{\rm LO}({\bf r},\mathbf{I},\bm{\sigma})
={}&-\frac{\nabla^2_{\bf r}}{2m_n}
+\frac{\mathbf {I}^2}{2\vartheta}
+\rm{C_0}\delta^{(3)}(r) 
-\beta P_2(\hat{\rm I}\cdot\hat r)\,
r^2\frac{d}{dr^2}\rm{C_0}\delta^{(3)}(r) \\
&+\rm{C_1}\,\bm{\sigma}\cdot \mathbf{I}\,\delta^{(3)}(r) 
-\beta P_2(\hat{\rm I}\cdot\hat r)\,
r^2\frac{d}{dr^2}\rm{C_1}\,
\bm{\sigma}\cdot\mathbf{I}\,\delta^{(3)}(r).
\end{split}
\label{eq:HLO}
\end{equation}
The first two terms are the one-body kinetic energies of the neutron and the rotor. The last four terms are rotor-neutron contact interactions, two of which feature a distorted contact operator. If hyperfine interactions are taken to be a leading-order effect, then the LO Hamiltonian for s-wave neutrons bound to the rotor contains three parameters: $\rm{C_0}$, $\beta$, and $\rm{C_1}$. 
Corrections to this Hamiltonian are suppressed by one power of $Q$ and/or two powers of $\xi$.

\subsection{Different regimes for the rotor and halo EFT expansions}

We now have two expansion parameters: $Q$ and $\xi^2$. In order to determine the combined power counting for operators in terms of $\xi$ and $Q$ we must specify the relative size of these parameters, which means we must specify the relationship between the energies associated with halo and rotor excitations. 

\subsubsection{$E_{\rm rotor} \ll E_{\rm halo}$}
For halo nuclei where the characteristic energy scale of the rotor $\rm{E_{rotor}}$ is small, but non-zero, compared to the energy scale of the halo, the rotor states form rotational bands, with the inter-band spacing governed by $\rm{E_{halo}}$. In this limit each rotational band can be treated according to the discussion in Ref.~\cite{ALCPP21}. The rotor motion can be computed in the adiabatic limit since the angular velocity of the valence neutron is much larger than the angular velocity of the core, leaving the valence neutron seeing a static core. The Hamiltonian can then be written in terms of one-body operators, that involve the overall angular momentum of the system and its projection on the $z$-axis in the intrinsic frame. 

In the extreme case where $\rm{E_{rotor}}=0$, the adiabatic approximation becomes perfect and all core states are degenerate. In that case the intrinsic Hamiltonian of the rotor $\rm{H_{rotor}^{LO}}$ in Eq.~(\ref{eq:H_tot}) can be neglected and the calculation reduces to the Nilsson model \cite{NILS55}. In this regime, it is efficient to solve the coupled-channel Schr\"odinger equations in the intrinsic frame of the rotor \cite{ESB00} and rotor excitation energies can be set to zero in those equations.

 \subsubsection{$E_{\rm rotor} \gg E_{\rm halo}$}
In the regime where the rotor energy scale lies well above the halo energy scale the decoupling theorem tells us that the theory may be recast as an EFT in which the rotor degree of freedom is integrated out of the theory. Deformation appears in such a theory only as a sub-leading effect. For example, the contribution of the quadrupole operator is of second order in the parameter $M_{\rm halo}/M_{\rm rotor}$\footnote{Here $M_{\rm rotor}=\sqrt{2 m_n E_{\rm rotor}}$.} and therefore can be treated perturbatively. At LO, the effective core-neutron potential $\rm{V_{cn}^{eff}}$ thus contains only the contact term (\ref{eq1}).
The fine-tuned bound-states described by such a Hamiltonian are $s$-wave states built on the $0^+$ core. The $2^+$ excitation can also be integrated out of the theory, thereby generating higher-body operators, see, e.g., Ref.~\cite{CPH20}. 
\subsubsection{$E_{\rm rotor} \sim E_{\rm halo}$}\label{subsec:protorgalphalo}
When the characteristic energies of halo and rotational excitations are of a similar size, the hierarchy between the different operators contributing to the effective potential $\rm{V_{cn}^{eff}}$ must be modified from the standard Halo EFT counting. In this regime, deformation becomes a leading-order effect, and the quadrupole operator (\ref{eq:Vhyp}) and the contact term (\ref{eq1}) both contribute to the halo-core interaction at leading order.

Under what typical conditions do these scales become similar? Since the core is a rigid ellipsoid at leading order we estimate 
\begin{equation}
\vartheta \sim \frac{1}{5} (\zeta^2 + \frac{1}{\zeta}) A m_n R_{\rm core}^2,
\end{equation}
where $A$ is the number of nucleons in the nucleus and $\zeta$ is the elongation parameter. Since we assumed ${\rm I} \sim 1$ we have $E_{\rm rotor} \sim E_{\rm halo}$ if
\begin{equation}
\frac{1}{2 m_n R_{\rm halo}^2} \sim \frac{5}{\zeta^2 A m_n R_{\rm core}^2} \quad \implies \quad Q^2 \sim \frac{10}{\zeta^2 A}.
\end{equation}
Thus the two energy scales are comparable if, for example, $Q \sim 1/2$, $\zeta \sim 2$, and $A = 10$, as is the case for ${}^{11}$Be. Systems which are more fine tuned, and thus have smaller $Q$'s, will achieve comparable scales with greater deformation, or if they occur further up the nuclear chart. 

In what follows we will apply this power counting to the system $^{11}$Be=$^{10}$Be($0^+,2^+$)+$n$ and $^{17}$C=$^{16}$C($0^+,2^+$)+$n$. To calculate leading-order predictions for these nuclei in this EFT requires the solution of coupled-channel Schr\"odinger equations for finite excitation energies. The argument of the previous paragraph implies that in these systems we can take $\xi \sim Q$ and unify the power counting. 

\section{Implementing the EFT at leading order}\label{impl_LO}
\subsection{The core-neutron potential}
\label{subsec:operators_powercounting}
Let us now establish the set of operators at leading order in our effective theory. To describe fined-tuned shallow $s$-wave bound states, following the Halo EFT prescriptions, at leading order, the core-neutron interaction is only a contact and, as such, its momentum-space representation is:
\begin{equation}
     {\rm{V_{ss}}}({\bf p}^\prime,{\bf p})=\rm{c_0} \quad,
     \label{eq:vLO}
\end{equation}
where $\rm{c_0}$ is a constant, proportional to $\rm{C_0}$, that is typically fitted to reproduce the binding energy or scattering length. 

Next-to-leading order terms enter at $O(Q)$. They are derivatives of the contact term, incorporated via:
\begin{equation}
    {\rm{V}}_{\rm NLO}({\bf p}^\prime,{\bf p})=\rm{c_0}+\rm{c_2(\textbf{p}^2+\textbf{p}^{\prime \, 2})} \quad,
    \label{eq:vNLO}
\end{equation}
where $\rm{c_2}$ is another coupling constant which is usually constrained using the Asymptotic Normalization Coefficient (ANC) or, equivalently, the effective range.

The core degrees of freedom associated with the rotational energy levels are coupled to the neutron co-ordinate via a quadrupole operator, i.e. a rank-2 tensor, of the following form 
\begin{equation} 
\rm{V_{sd}({\bf p}^\prime,{\bf p})=c_{\rm{sd}} \big [\mathbf{I} \cdot \textbf{q} \cdot \mathbf{I} \cdot \textbf{q} -\frac{1}{3} \mathbf{I}^2\textbf{q}^2 ]} \hspace{1cm}
\text{with} \hspace{0.3cm}\textbf{q}=\textbf{p}-\textbf{p}' \quad,
\label{eq:quadop}
\end{equation}
where $\rm{\mathbf{I}}$ is the spin of the core and $\rm{\mathbf{q}}$ is the momentum conjugate to the neutron co-ordinate $\rm{\mathbf{r}}$. The associated coupling constant $\rm{c_{sd}} (\propto \rm{C_{sd})}$ can be related to the usual geometrical deformation parameter $\beta_2$ in the rotational model. This coupling constant can be fitted to the spectrum of the halo nucleus. Note that this operator, like Eqs~(\ref{eq:vLO}) and (\ref{eq:vNLO}), is defined in momentum space, although we solve the entire problem in coordinate space, as will be shown later.

The leading-order potential is completed by a hyperfine-like operator describing the interaction between the spin of the core $\rm{\mathbf{I}}$ and the spin of the valence nucleon $\rm{\mathbf{j}}$, for which we adopt the form
\begin{equation}
 \rm{V_{hf}({\bf p}',{\bf p})=c_{hf} \mathbf{I} \cdot \mathbf{j}} \quad,
 \label{eq:Vhyp}
\end{equation}
where $\rm{\mathbf{j=\boldsymbol{\ell} + s}}$, $\boldsymbol{\ell}$ being the relative orbital angular momentum of the valence neutron and $\mathbf{s}$ the spin of the valence nucleon. This operator equals the leading-order operator (\ref{eq:HF}) for $s$-waves, so replacing $\bm{\sigma}/2$ by $\mathbf{j}$ is a higher-order effect. At leading order, the hyperfine-like operator~(\ref{eq:Vhyp}) is actually not needed to renormalize binding energies but is introduced for phenomenological reasons. In its absence the $\frac{5}{2}^+$ and $\frac{3}{2}^+$ states would collapse into one degenerate state. We must also add the effects of deformation on the potential (\ref{eq:Vhyp}), so we also include the term
\begin{equation}
\rm{V_{hf,def}({\bf p}',{\bf p})= c_{hf}\big [{\mathbf{I}} \cdot \textbf{q} \cdot {\mathbf{I}} \cdot \textbf{q} -\frac{1}{3} \mathbf{I}}^2 \textbf{q}^2\big](\mathbf{I} \cdot \mathbf{j}) \quad .
\label{eq:Vhypdef}
\end{equation}
In practice, when implementing Eq.~(\ref{eq:Vhypdef}) we choose the symmetric combination of {\bf I} operators in going from the classical picture of the deformation operator to a quantum formulation. We implement this by computing the off-diagonal channels induced by the operator (\ref{eq:Vhypdef}) as the average of the contributing strengths from the two different operator orderings. Note that at LO the strength of this distorted hyperfine potential is not a free parameter, since it is determined by ${\rm c_{hf}}$ and the amount of deformation. 

At leading order, the effective core-neutron potential $\rm{V^{eff}_{cn}}$ is then the sum of the operators in Eqs.~(\ref{eq:vLO}),  (\ref{eq:quadop}),  (\ref{eq:Vhyp}), and (\ref{eq:Vhypdef}). There are three coupling constants $\rm{c_0}$,$\rm{c_{sd}}$ and $\rm{c_{hf}}$ that must be fitted to reproduce the properties of the spectrum of the halo nucleus of interest. After Fourier Transforming to co-ordinate space, the effective $\rm{LO}$ core-neutron Hamiltonian describing the deformed halo nucleus can be written as
\begin{equation}  
\rm{ \hat{H}_{cn}^{LO}(\textbf{r},\eta})=
\hat{T}_{\textbf{r}}
 +V_{cn}^{eff}(\textbf{r},\eta) + H_{rotor}^{LO}(\eta) \quad,
 \label{eq:H_tot}
\end{equation}
where $\rm{V_{cn}^{eff}(\textbf{r},\eta)}=V_{ss}(r)+V_{sd}({\bf r},\eta) +V_{hf} (r)+ V_{hf,def}({\bf r},\eta)$. The orientation of the non-spherical core is parameterized here using the three Euler angles $\eta$ which define the three-dimensional rotation $\eta \equiv (\alpha,\beta,\gamma)$, transforming the laboratory frame into the intrinsic frame. The effective potential $\rm{V_{cn}^{eff}}$ is not central, and depends on the core's orientation. We then define 
\begin{equation}
\cos \theta=\frac{\rm{\hat{\bf I} \cdot \hat{r}}}{I}
\end{equation}
as the angle between the core spin and the neutron co-ordinate ${\bf r}$. The leading contribution~\footnote{Leading both in powers of $\beta$ and in the EFT expansion.} to each of $\rm{V_{sd}}$ and $\rm{V_{hf,def}}$ is then the piece of Eq.~(\ref{eq:deformedcore}) that is proportional to $P_2(\cos \theta)$. 

 To perform these calculations, following Ref.~\cite{CAP18}, the contact and hyperfine interactions (\ref{eq:vLO}) and (\ref{eq:Vhyp}) are regularized using a Gaussian of width $\sigma$ so they become
\begin{equation}
 \rm{V_{ss}(r;\sigma)=C_0 e^{-\frac{r^2}{2\sigma^2}}}; \quad  \rm{V_{hf}(r;\sigma)=C_{hf} e^{-\frac{r^2}{2\sigma^2}}} {\bf I} \cdot {\bf j} ,
    \label{eq:vLOreg}
\end{equation}
where $\sigma$ is a parameter that defines the dividing line between short-range physics that has been integrated out and the long-range physics that is being treated explicitly. (Meanwhile, the coupling constants $\rm{C_0}$ and $\rm{C_{hf}}$ are the analogs in co-ordinate space of the lowercase constants $\rm{c_0}$ and $\rm{c_{hf}}$ defined in momentum space in Eqs.~(\ref{eq:vLO}) and (\ref{eq:Vhyp}) respectively.) We choose the Gaussian regulator partly because it leads to local potentials at leading order.
For this regulator the $sd$ mixing potential is:
\begin{equation}
\rm{{V}_{sd}}({\bf r},\eta)=\rm{C_{sd}} P_2(\cos \theta) \frac{r^2}{\sigma^2} e^{-\frac{r^2}{2\sigma^2}},
\label{eq:VsdGaussian}
\end{equation}
with $\rm{C_{sd}} \equiv \frac{1}{2} C_0 \beta$. We note that this is the same as adopting:
\begin{equation}
\rm{V_{sd}}({\bf r},\eta)=\beta P_2(\cos \theta) \sigma \rm{\frac{d}{d \sigma}} \rm{V_{ss}(r;\sigma)} \quad,
\end{equation}
based on Refs.~\cite{KC25,KC26} and previous calculations of neutron-core systems that use the particle-rotor model \cite{ESB95,NUNES96,THOM04}. The implementation of $\rm{V_{hf,def}}$ is analogous. Equations (\ref{eq:H_tot})--(\ref{eq:VsdGaussian}) then define a local, Gaussian-regulated version of the leading-order deformed Halo EFT Hamiltonian (\ref{eq:HLO}). 

By performing calculations for different values of $\sigma$ (in our case: $\sigma=1.3, 1.5$ and $2.0$~fm) we can assess the sensitivity of our results to short-range physics at leading order in the EFT.

\subsection{Solving for EFT energy levels}
To solve numerically the eigenvalue problem satisfied by the effective core-neutron Hamiltonian, we expand it in a convenient basis constructed as follows. The intrinsic Hamiltonian $\rm{\hat{H}_{rotor}^{LO}}$ of the core is characterized by intrinsic core states $\rm{\phi_{M_{c}}^{I_{c}^{\pi_c}}}$ satisfying the eigenvalue equation
\begin{equation}
   \rm{H_{rotor}^{LO}\phi_{M_{c}}^{I_{c}^{\pi_c}}(\eta)=\epsilon_{c}^{I_c^{\pi_c}} \phi_{M_{c}}^{I_{c}^{\pi_c}}(\eta)} \quad,
   \label{eq:coreeigeq}
\end{equation}
where $\rm{\epsilon_c}$ are the core excitation energies, $\rm{\pi_c}$ is the parity of the core, $\rm{I_c}$ is its spin and $\rm{M_c}$ is the spin projection. In our approach, we consider a rotational model for the core \cite{BOMOT69,ESB95,NUN96,THOM04}, where it is described as an axially symmetric rigid rotor. The rotor eigenstates $ \phi_{M_{c}}^{I_{c}\pi_c}$ are expressed as a linear combination of the Wigner matrices ${\cal D}^{I_c}_{M_cK_c}$ \cite{BOMOT69}
\begin{equation}
    \rm{\phi_{M_{c}}^{I_{c}\pi_c}({\eta})=\sqrt{\frac{2I_c+1}{16\pi^2}}i^{\frac{1-\pi_c}{2}}
 \left[{\cal D}^{I_c}_{M_cK_c}({\eta})+\pi_c(-1)^{I_c+K_c}{\cal D}^{I_c}_{M_c\,-K_c}({\eta})\right]} \quad,
 \label{rotoreigstat}
\end{equation}
with $\rm{K_c}$ the projection of the spin in the intrinsic rest frame of the core. As this is a constant value within a given rotational band, we ignore it hence forward in the notation of the core eigenstates~(\ref{eq:coreeigeq}).
Each state $\rm{\Psi^{J^\pi M}}$ of the $c$-$n$ spectrum is defined by its total angular momentum $\rm{J}$ and parity $\rm{\pi}$, $\rm{M}$ denoting the projection of $\rm{J}$. These states can be expanded in a basis separating the intrinsic structure of the core from the $c$-$n$ relative motion
\begin{equation}
   \rm{ \Psi^{J^\pi M}(\textbf{r},\eta)=\sum_{\alpha}} i^\ell \rm{\frac{u_{\alpha}(r)}{r} [\mathcal{Y}_{\ell j}(\hat{r}) \otimes \phi_{M_{c}}^{I_{c}\pi_c}(\eta)]^{J M}}\quad,
    \label{eq:totalwf}
\end{equation}
where $\alpha$=$\rm{\left\{\ell,s,j,I_c,\pi_c\right\}}$ defines a channel, i.e., a set of quantum numbers describing each allowed core-neutron configuration. By substituting Eq.~(\ref{eq:totalwf}) into the Schr\"odinger equation for the $c-n$ Hamiltonian (\ref{eq:H_tot}), we obtain a set of coupled-channel Schr\"odinger equations
\begin{equation}
    \left[
    \rm{T_r^{\ell}}
    +V_{\alpha \alpha}(r) + \epsilon_{\alpha}- E\right] \psi_{\alpha}(\rm{r}) =-\sum_{\alpha'\neq \alpha}\rm{V_{\alpha \alpha'}(r)}\psi_{\alpha'}(r) \quad,
    \label{eq:coupledeSCH}
\end{equation}
with the matrix elements $\rm{V_{\alpha \alpha'}(r)}$ = $\langle \mathcal{Y}_{\alpha}(\rm{\hat{r}})  \chi_{\alpha}(\eta)|{\rm{V_{cn}^{eff}(\textbf{r},\eta)}}|{\mathcal{Y}_{\alpha'}(\rm{\hat{r}}) \chi_{\alpha'}(\eta)} \rangle$ governing the coupling the different channels. We solve these coupled-channel equations using the R-matrix method on a Lagrange mesh, which allows us to treat bound and scattering states on the same footing \cite{HES98,Desc10,BAY15}. For bound states, wave functions and ANCs are computed for each channel $\alpha$. For continuum states, the collision matrix, and thus the scattering phaseshifts, and scattering wave functions are also calculated. In particular, for resonant states, we use an iterative version of the R-matrix method introduced by Schneider and then Descouvemont and Vincke \cite{SCHNEI81,DESCVIN90}. It allows us to iteratively compute resonant properties, i.e. their energies and widths, from the resolution of a non-Hermitian eigenvalue problem. A recent review on the subject is available in Ref.~\cite{DESCDOH24}.
\subsection{Coulomb dissociation}

With the positive-parity spectrum and wave functions of a rotor-halo in hand, in particular its ground state, we can compute the reduced E1 transition probability $\rm{dB(E1)/dE}$. This observable corresponds to the response of the $c$+$n$ system to an electric dipole stimulus, leading to its breakup. This enables us to access the nuclear structure of the nucleus as the E1 transition occurs from the initial bound ground state to some final continuum state of the system. Using semi-classical arguments, the $\rm{dB(E1)/dE}$ can be related to the Coulomb dissociation cross section and can therefore be extracted from experimental data for comparison~\cite{ALDWIN75,BERBAU88}. Following Typel and Baur \cite{TYPBAU05}, the E1 reduced transition probability between states of total angular momentum $\rm{J_i}$ and $\rm{J_f}$ can be computed as

\begin{equation}
\label{beTYPBAU}
\rm{\frac{d{\cal B}(E\lambda, J_i \rightarrow J_f)}{dE}= \frac{2 J_f +1 }{2
  J_i+1} \frac{\mu k }{(2 \pi)^3 \hbar^2} \sum_{j_f \ell_f} \left | \sum_{j_i \ell_i I_c}  \langle k J_f j_f \ell_f s I_c|| \mathcal{M}(E\lambda) || J_i j_i \ell_i s I_c \rangle  \right |^2 } \nonumber\quad,
\end{equation} 
where the multipole operator is taken from Ref.~\cite{ARA10}, $\rm{k=\sqrt{2\mu E/\hbar^2}}$ and $\rm{E}$ is the core-neutron relative energy. For the continuum states, the convention $\rm{\langle k J
| k' J \rangle = \delta(k-k')}$ is used and their asymptotic behaviour is given by 
\begin{eqnarray}
\label{eq:wfasympt}
\rm{u_{\alpha'}(r)=  
 \frac{i }{2} 
 \Big[ I_{\alpha}(k_{\alpha} r)\delta_{\alpha' \alpha} -\left( \frac{k_{\alpha}}{k_{\alpha'}}\right)^{\frac{1}{2}}
S_{\alpha' \alpha}O_{\alpha'}(k_{\alpha'}r) \Big] } \quad,
\end{eqnarray}
where $\alpha$ designates the entrance channel and $\rm{k_{\alpha}=\sqrt{2\mu \abs{(E-\epsilon_{\alpha})}/\hbar^2}}$. Here, $\rm{I_{\alpha}}$ and $\rm{O_{\alpha}}$ denote the incoming and outgoing Coulomb wave functions as defined in Ref.~\cite{DESC2016}, whereas $\rm{S_{\alpha' \alpha}}$ is the scattering matrix of the system. At first order of perturbation theory, using the equivalent photon method \cite{ALDWIN75,BERBAU88}, the $\rm{dB(E1)/dE}$ calculated using Eq.~(\ref{beTYPBAU}) can be related to the differential Coulomb dissociation cross section $\rm{d\sigma/dE}$ through the following expression
\begin{equation}\label{eq:dsigdE}
\rm{\frac{d\sigma}{dE}=\frac{16\pi^3}{9\hbar c}N_{E1}(E)\frac{dB(E1)}{dE}} \quad,
\end{equation}
where $\rm{N_{E1}}$ is the number of equivalent photons of energy $\rm{E}$, which is exchanged between a high velocity projectile passing through the Coulomb field generated by a target of high nuclear charge $Z$. A simple closed expression for $\rm{N_{E1}}$ is given in Ref.~\cite{BERBAU88}. 
\\

This reaction model assumes that the reaction is only driven by the Coulomb interaction between the core and the high-Z target, thus completely neglecting the contribution of the nuclear interaction, which affects the reaction process \cite{TS01,CBM03c}. Higher-order effects are also neglected and do need to be taken into account \cite{EBS05,CB05}. Although limited, this direct E1 transition reaction model reproduces forward-angle data fairly well, hence its use in the next section \cite{CN17}. Nevertheless, full reaction calculations would require to couple our EFT description of deformed halo nuclei to a full reaction model, which incorporates both the nuclear contribution and higher-order effects, such as the dynamical eikonal approximation~\cite{DEA1,DEA2}.

\section{Results}\label{sec:cases}
We apply our effective theory to $^{11}$Be, the archetypal halo nucleus, and $^{17}$C, both of which have been calculated in multiple models \cite{ESB95,VINHMAU95,NUN96,CDN10,CRES11,MORCRES12,CAP22,DEL23,TIMO10,AMOS12,SMA15,KIM23,CHDESC23,PUN23,PUN25,SUH25,PUN26}. Both are characterized by at least one shallow $s$-wave bound state, and also exhibit a core with a suspected or confirmed rotational-band structure: a $0^+$ bandhead and a $2_1^+$ first excited state \cite{KHD99,McCoy24}. We first show that it is possible to reproduce the low-lying positive parity states of the $^{11}$Be and $^{17}$C $c$-$n$ spectrum using the defined leading-order operators in our EFT (see Sec.~\ref{subsec:operators_powercounting}). Secondly, using the calculated spectrum, we compute Coulomb breakup observables and compare our calculations to experimental data.

We compute the positive parity states of $^{11}$Be and $^{17}$C, up to the core($2_1^+$)+$n$ threshold, by fitting the coupling constants $\rm{C_0}$, $\rm{C_{sd}}$, and $\rm{C_{hf}}$ to their low-energy spectrum. The procedure by which we fit these constants is as follows. 

First, we identify a shallow $s$-wave bound state, i.e. $\frac{1}{2}^+$, in their low-energy spectrum and we tune $\rm{C_0}$ to reproduce its experimental one-neutron separation energy. If the contact operator multiplying $\rm{C_0}$ were the only interaction in the Hamiltonian the low-lying $\frac{5}{2}^+$, $\frac{3}{2}^+$ would be degenerate, lying in the $c-n$ continuum at the energy $\rm{E=-S_n+E_{2^+}}$. The quadrupole operator, and associated LEC $\rm{C_{sd}}$ is then introduced, allowing coupling between $s$ and $d$-waves and the $0^+$ and $2^+$ states of the core. This operator can be used to adjust the position of the $\frac{3}{2}^+$-$\frac{5}{2}^+$ doublet relative to the $\frac{1}{2}^+$ state. The 
hyperfine strength $\rm{C_{hf}}$ is then tuned to reproduce the experimental splitting $\rm{\Delta_{3/2^+-5/2+}}$ between the $\frac{3}{2}^+$ and $\frac{5}{2}^+$ states. The $\frac{3}{2}^+ - \frac{5}{2}^+$ experimental splitting in these systems $\rm{\Delta_{3/2^+-5/2+}}$ is of order ${\rm E_{rotor}=E_{2_1}^+}$ \cite{AJZ90,KELL12,ELE05,BOHL07,TILL93}. 
\subsection{$^{11}$Be}
The low-energy spectra of $^{11}$Be and $^{10}$Be are experimentally well documented. The $^{10}$Be core exhibits a $0^+$ ground state and a first excited $2_1^+$ state 3.368~MeV above it \cite{AJZ90}. The $0^+$ state can be seen as the bandhead of a ground-state rotational band of which $2^+_1$ is a part \cite{KHD99,McCoy24}. The $^{11}$Be halo nucleus is characterized by a $\frac{1}{2}^+$ ground state with a low separation energy $\rm{S_{1n}}=0.5$~MeV \cite{KELL12}. The ANC of the $\frac{1}{2}^+$ bound state of $^{11}$Be is also known from \textit{ab initio} calculations \cite{CAL16}. In our deformed core halo EFT one high-energy scale is set by the first non-rotational excitation in the $^{10}$Be core's spectrum, which is a 2$^+_2$ state lying at 5.96~MeV \cite{AJZ88}. The high-energy scale can be estimated from the size of the $^{10}$Be core. Different experiments and theoretical calculations yield a $^{10}$Be root-mean-square matter radius of 2.2--2.3~fm \cite{ARA04,OGA00,ALKH96,LI25,TAN88}.

In our EFT, two different expansion parameters $Q$ and $\xi$ appear, coming from expansions of two distinct Lagrangians, namely the halo EFT and rotor Lagrangians described in Sec.~\ref{sec:haloeft} and \ref{sec:rotoreft}. In Sec.~\ref{subsec:protorgalphalo}, we discuss the regime $\rm{E_{rotor} \sim E_{halo}}$, which is of interest here. In this regime, the rotor and halo scales are of the same order of magnitude, leading us to study a regime where deformation enters at leading order. Nevertheless, this does not mean that the expansion parameters $Q$ and $\xi$ are identical.

Firstly, the halo EFT Lagrangian is expanded in powers of the small parameter $Q$, ratio of distance (mass) scales associated with the core and the halo. This Lagrangian describes the physics of halo nuclei, i.e. systems dominated by the asymptotic part of their wavefunction. For $^{11}$Be, the ratio between the momentum scales of the halo and the rotor is moderate, $Q=\rm{\sqrt{2 m_n S_{1n}} R_{core}} \sim 0.4$ \cite{HAM17}.

At leading order, only the $s$-waves are taken into account. When computing the E1 dissociation of ${}^{11}$Be, $p$-wave interactions are a next-to-leading-order effect~\cite{HP11}. At that same order a two-derivative $s$-wave contact operator appears in the Lagrangian. Once these terms are included the impact of the remaining terms becomes of order $O(Q^2)$. Thus, higher-order finite range effects can be systematically incorporated, by writing relevant operators including increasing powers of the expansion parameter $Q$.
In this section, we only perform calculations for $^{11}$Be, which are leading order in terms of the expansion parameter $Q$. 

Secondly, we incorporate rotational core degrees of freedom in our EFT using a rotor-Lagrangian. Starting from a LO rigid-rotor picture as in Eq.~(\ref{eq:LOHrotor}), the EFT of nuclear rotations allows us to organize corrections to the rigid-rotor in powers of the expansion parameter $\xi$, ratio between the angular velocity of the probed rotor scale over the angular velocity associated with the first non-rotational excitation of the core, which is the high scale. 

At leading order, Eq.~(\ref{eq:LOHrotor}) denotes the intrinsic Hamiltonian of the rigid rotor. Omitted operators are suppressed by $\rm{(v/v_{hi})^2}$, i.e. by $\rm{E_{2^+_1}/E_{hi}}$. For the $^{10}$Be core, the rotor expansion parameter $\xi^2$ can be estimated as $\rm{E_{2_1^+}/E_{break}}\sim0.56$, 
where $\rm{E_{2_1^+}}$ is determined by the first excited state of the $^{10}$Be core and $\rm{E_{break}}$ is set by the energy of its first $2_2^+$ non-rotational state. In this manuscript, we consider only leading-order effects in the rotor-Hamiltonian. Higher-order operators including, for example, non-rigidity effects, are not taken into account and will be studied in a separate manuscript in the future.  
The obtained expansion parameter is quite large. This may lead to a rather slow convergence of the rotor-EFT. Significant subleading corrections are thus expected but will not be investigated in this manuscript.

We calculate the positive parity states of the $^{11}$Be spectrum up to the $^{10}$Be($2_1^+$)+$n$ threshold for three different regulator parameters. The $\frac{1}{2}^+$ ground state is described as a state mixing different configurations of the core, which could be in either its $0^+$ or $2_1^+$ state, with the neutron being bound correspondingly in an s-wave or d-wave orbital. 
The calculated energy levels are compared to the experimental energy levels and displayed in Tab.~\ref{tab:11Be_nrjlevels}, along with the widths of the $\frac{5}{2}^+$ and $\frac{3}{2}^+$ resonant states. At leading order, the resonance widths are predictions of our effective theory. 

In Tab.~\ref{tab:11Be_nrjlevels}, for each regulator parameter $\sigma$, the experimental energy levels are reproduced at the correct position. The set of defined leading-order operators is sufficient to renormalize binding energies. However, we recall that the hyperfine operator~(\ref{eq:Vhyp}) is not needed to renormalize binding energies at leading order, but is promoted to LO for phenomenological reasons. The obtained ANCs are mildly regulator dependent: their variation of $\pm 20$\% is consistent with the expected size of next-to-leading-order effects. The ANCs are in fair to good agreement with the {\it ab initio} ANC, with the level of agreement dependent on the value of the regulator parameter $\sigma$. 

The predicted LO resonance widths of the $\frac{5}{2}^+$ and $\frac{3}{2}^+$ states are also displayed in Tab.~\ref{tab:11Be_nrjlevels}. These widths are clearly regulator-dependent. While the width of the $\frac{3}{2}^+$ state increases monotonically with the regulator $\sigma$, this is not the case for the $\frac{5}{2}^+$ width, which has a more complicated behaviour. These results stem from the interplay between the two allowed decay mechanisms, namely coupling via the deformed contact and hyperfine contact operators. 

Both resonances are known to be quite narrow \cite{KELL12}. The EFT successfully predicts that the $\frac{5}{2}^+$ is narrow, but it always predicts a central value for the $\frac{3}{2}^+$ width that is too large. However, in the latter case the LO result given in Table~\ref{tab:11Be_nrjlevels} has uncertainties which are sufficiently big that they still cover the experimental value. In contrast, for two out of the three $\sigma$ values examined, the $\frac{5}{2}^+$ state is predicted to be so narrow that the fractional uncertainties we have assigned do not encompass the experimental result.
This difference in the behavior of the LO EFT predictions is likely connected to the fact, pointed out in Refs.~\cite{LAY12,CAP18,CAP22,PUN23,PUN26}, that these two resonant states couple differently to the first $2^+$ excited state of the $^{10}$Be. The small widths and cutoff sensitivity seen for both resonance widths imply that the Halo EFT prediction for them should be understood as being consistent with zero at leading order. The renormalization of these widths will occur at higher orders in the EFT, through operators that encode short-distance structure effects like those discussed in Refs.~\cite{LAY12,CAP18,CAP22,PUN23,PUN26}.

\begin{figure}[hb]
    \centering
    \begin{minipage}{0.48\linewidth}
        \centering
        \includegraphics[width=\linewidth]
        {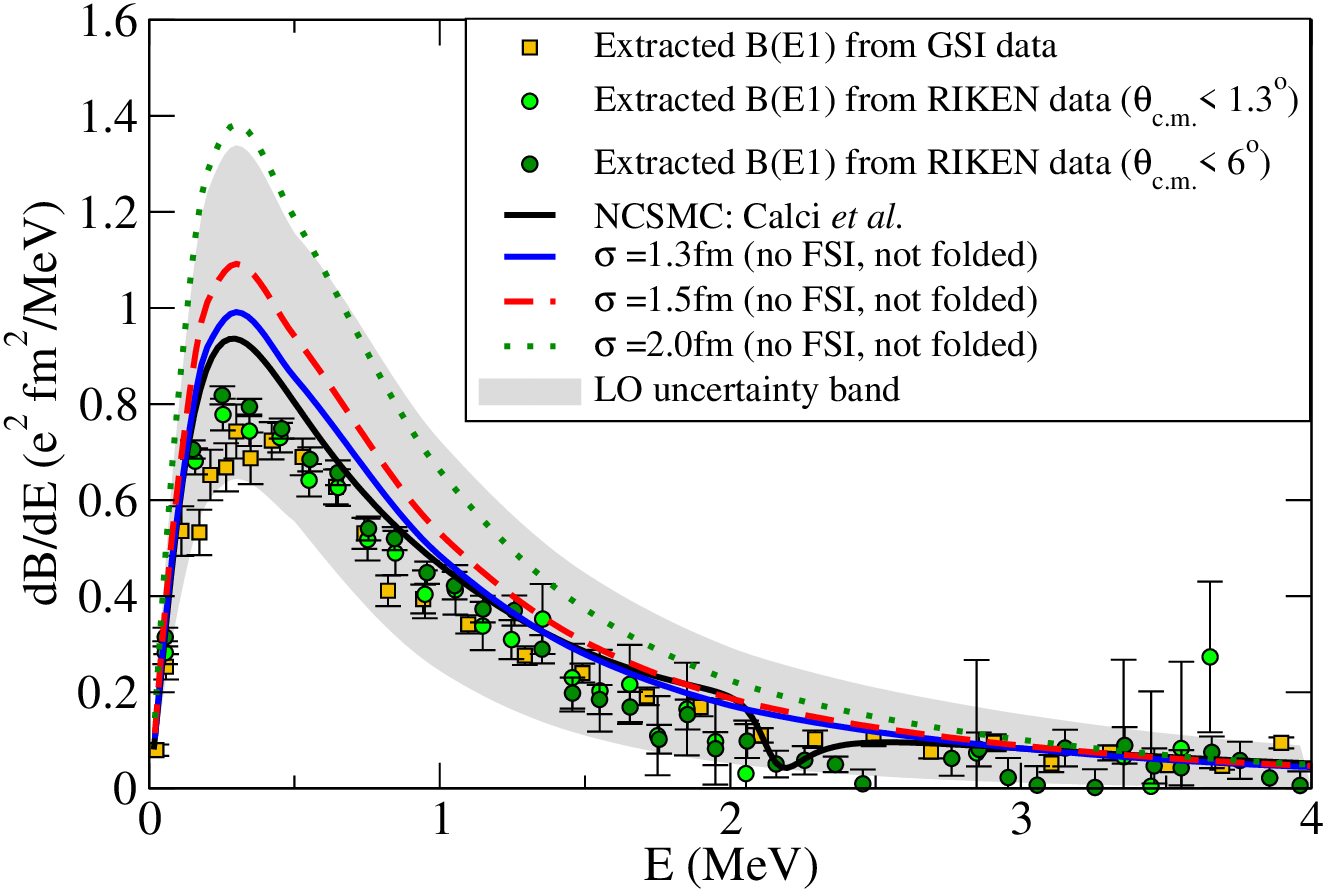}
    \end{minipage}\hfill
    \begin{minipage}{0.48\linewidth}
        \centering
        {\includegraphics[width=\linewidth]
        {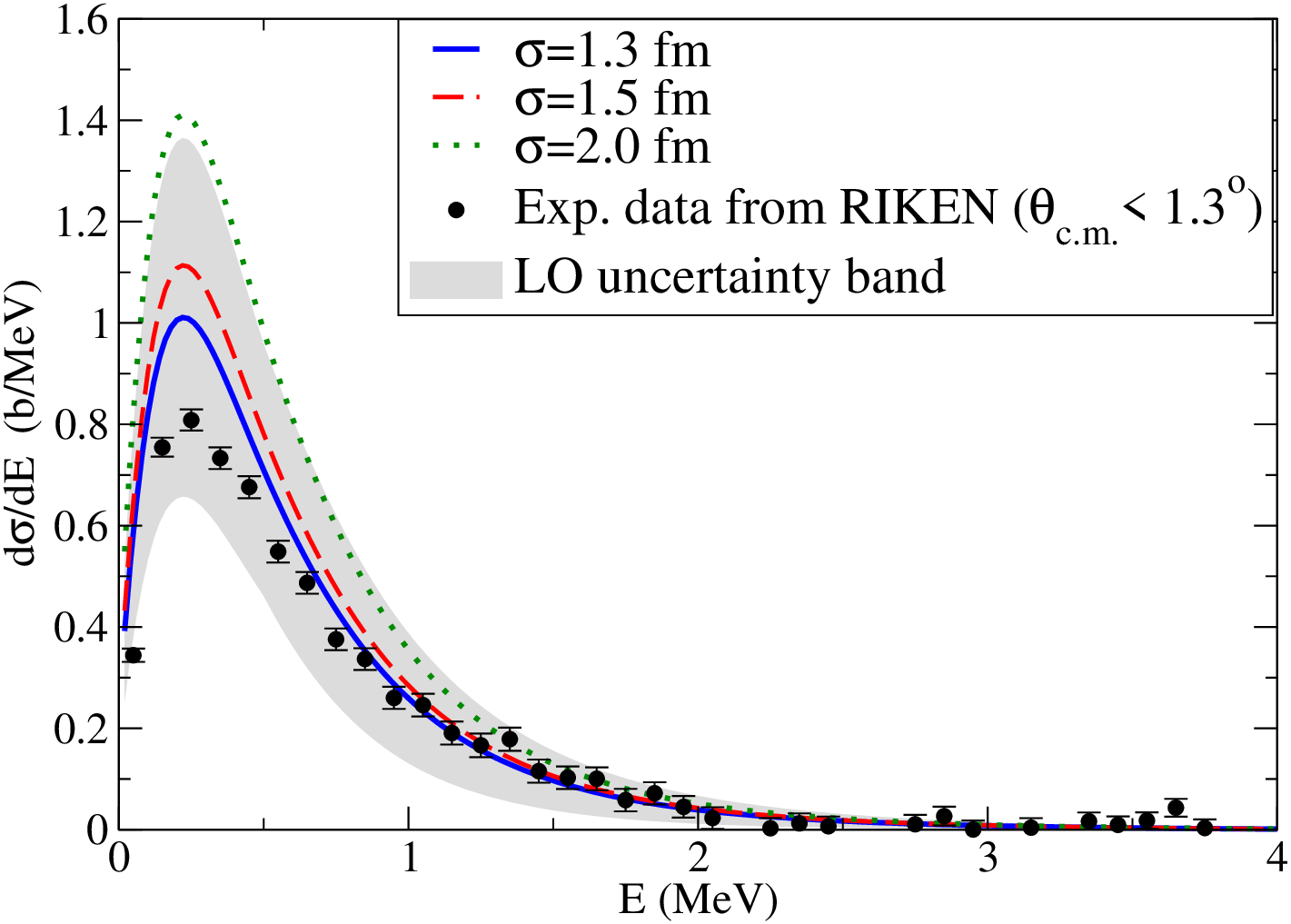}}
    \end{minipage}
    \caption{Leading order calculations of the $\rm{B(E1)}$ distributions (top panel) and the Coulomb breakup cross section (bottom panel) of $^{11}$Be: $\rm{^{11}Be + Pb\rightarrow}^{10}Be+n+Pb$ at 69~MeV per nucleon. Our calculations are performed without final-state interaction in the $p$-waves and the breakup cross section is folded with the experimental resolution. Experimental data from RIKEN~\cite{FUKU04} and GSI~\cite{PAL2003} are also displayed. \textit{Ab initio} NCSMC calculation of Ref.~\cite{CAL16} are shown for comparison.}
    \label{fig:dbe1fig11Be_LO}
\end{figure}

\begin{table}[ht]
\caption{\label{tab:11Be_nrjlevels} Table summarizing the properties of positive parity states of $^{11}$Be, at leading order, up to the $^{10}$Be($2_1^+$)-$n$ threshold. The energies $\rm{E}$ and widths $\Gamma$ are in MeV and 
the asymptotic normalization coefficients ${\cal C}$ are given in $\rm{fm}^{-\frac{1}{2}}$. The ANC of the $\frac{1}{2}^+$ state of $^{11}$Be has also been computed from \textit{ab initio} No-Core Shell Model with Continuum calculations (NCSMC) and is shown~\cite{CAL16}. Experimental data are also displayed for comparison~\cite{KELL12}. The inputs for the EFT calculations are shown in bold, while the predictions are displayed in regular text with their corresponding LO uncertainties.}
\begin{ruledtabular}
\begin{tabular}{cccccccc}
 $\sigma$ (fm) & 1.3 & 1.5 & 2.0 & $\rm{Exp.}$\cite{KELL12} & $\rm{NCSMC}$ \cite{CAL16} \\ \hline \hline 
$\rm{E_{1/2^+}}$ & $\mathbf{-0.5031}$ & $\mathbf{-0.5031}$ & $\mathbf{-0.5031}$  & $-0.5016$ & $-0.50$\\
${\cal C}_{1/2^+}$ & $0.75 \pm 0.26$ & $0.79 \pm 0.28$ & $0.89 \pm 0.31$ & ...  & $0.786$ \\\hline
$\rm{E_{5/2^+}}$ & $\mathbf{1.274}$ & $\mathbf{1.274}$ & $\mathbf{1.274}$ & $1.274$  &  $1.31$ \\
$\Gamma_{5/2^+}$ & $(3.60 \pm 2.00)\times 10^{-3}$ & $0.11 \pm 0.06$ & $(2.41 \pm 1.35)\times 10^{-2}$  & $0.100$ & $0.10$ \\ \hline
$\rm{E_{3/2^+}}$ & $\mathbf{2.90}$ & $\mathbf{2.90}$ & $\mathbf{2.90}$  & $2.90$  &   $2.92$\\
$\Gamma_{3/2^+}$ & $0.26 \pm 0.22$ & $0.55 \pm 0.46$ & $0.97 \pm 0.82$  &  $0.122$ &  0.06\\ 
\end{tabular}
\end{ruledtabular}
\end{table}
Turning to the $\rm{B(E1)}$ strength, in Halo EFT, $p$-wave final-state interactions are next-to-leading order in the continuum~\cite{HP11}. In this LO calculation we therefore describe the $p$-wave scattering states of $^{10}$Be and the neutron as plane waves. The results that are then obtained for the differential $\rm{B(E1)}$ distribution (corresponding Coulomb cross section)  are displayed for different regulators, in the top (bottom) panel of Fig.~\ref{fig:dbe1fig11Be_LO}. Our $\rm{B(E1)}$ distributions are not folded with the experimental resolution, whereas our Coulomb cross sections are, before being compared with RIKEN experimental data \cite{FUKU04}.  In the top panel of Fig.~\ref{fig:dbe1fig11Be_LO}, it is clearly visible that the leading order $\rm{B(E1)}$ calculations are in fair to good agreement with the experimental data---depending on the regulator. The maximum of the distributions occurs, as expected, at $\rm{E}=\frac{3}{5}\rm{S_n}$ \cite{FUKU04}.  Correspondingly, for the Coulomb cross sections (see bottom panel of Fig.~\ref{fig:dbe1fig11Be_LO}), our LO calculations are also in fair to good agreement with the data depending on the regulator, with the high-energy part of the cross section being well reproduced and the low-energy one presenting an overshoot around the peak. 

Figure~\ref{fig:dbe1fig11Be_LO} shows the B(E1) distributions together with the Coulomb breakup cross sections. The grey bands represent the LO uncertainty, and provide estimates of the size of the omitted higher-order effects. We estimate the corresponding relative uncertainties using the prescription $\max(k,\gamma)/M_{\mathrm{hi}}$, where $k=\sqrt{2m_nE}$ and $\gamma=\sqrt{2m_nS_{1n}}$ denote the scattering and binding momenta, respectively. We observe that our results obtained with the different regulators mostly lie within these uncertainty bands, supporting our simple estimate of the theoretical uncertainties. Similarly, most of the experimental data points lie within the uncertainty bands. This suggests that the discrepancy between our LO results and the experimental data can be accounted for by incorporating higher-order operators omitted at LO.

As mentioned above, at leading order, only the binding energy of the $\frac{1}{2}^+$ ground state of $^{11}$Be is reproduced, while the ANC is not and has some dependence on the regulator (see Table~\ref{tab:11Be_nrjlevels}). The $\rm{dB(E1)/dE}$ is a peripheral observable which strongly depends on the ANC and therefore reflects this dependence on the regulator. The ANC could be renormalized at NLO by introducing a two-derivative operator (\ref{eq:vNLO}) involving derivatives of the $s$-wave contact term. This will be discussed in Sec.~\ref{sec:beyondlo}, when we discuss beyond-leading-order effects. Note also that the dip observed at 2.15~MeV in the \textit{ab initio} $\rm{B(E1)}$ prediction is not reproduced by any of our positive-parity states calculations. This was expected since this energy corresponds to the presence of a $\frac{3}{2}^-$ resonance,  which is not described as our LO calculations do not include final-state interactions. 
\subsection{$^{17}$C}
The low-energy spectrum of $^{17}$C has been studied by multiple experimental measurements. It has been established that its ground state is a $\frac{3}{2}^+$ bound by 0.73~MeV dominated by a  $^ {16}$C(2$^+$)+$n$ configuration \cite{DATTA03}. It also exhibits two other shallow bound states $\frac{1}{2}^+$(-0.52~MeV) and $\frac{5}{2}^+$ (-0.40~MeV) \cite{ELE05,BOHL07}. Two narrow $\frac{7}{2}^+$ and $\frac{9}{2}^+$ resonances have also been identified in the $^{17}$C spectrum \cite{SAT08}. The $^{16}$C core also features a rotational band structure, with a $0^+$ ground state and a first excited $2_1^+$ state at 1.766~MeV \cite{TILL93}. The first non-rotational excitation in the spectrum of the core is located at 3.03~MeV \cite{TILL93}. It sets one high-energy scale of the problem. Another high-energy scale can be related to the size of the $^{16}$C core. The matter radius of $^{16}$C has been recently measured and is about 2.8~fm~\cite{KAN16}. 

As for $^{11}$Be, we treat $^{17}$C in the regime where $\rm{E_{rotor} \sim E_{halo}}$ and consider the two different expansion parameters $Q$ and $\xi$. In this picture, deformation is a leading-order effect. In this section, we consider only LO effects in both the halo EFT expansion parameter $Q$ and the rotor one $\xi^2$. From halo EFT, we only include (deformed) $s$-wave $c$-$n$ contact interactions, as in Eq.~(\ref{eq:vLO}). 

On the one hand, the $\frac{1}{2}^+$ state is a halo state \cite{CHDESC23} bound by only 0.52~MeV. Considering the experimental separation energy of that state and the size of the ${}^{16}$C core, the expansion parameter governing finite-range effects is again found to be moderate to relatively large: $\rm{Q=\sqrt{2 m_n S_{1n}} R_{{}^{16}C}}
\approx 0.43$.

On the other hand, the expansion parameter $\xi^2$ governing the leading-order rigid rotor Hamiltonian~(\ref{eq:vLO}) is accurate to $\rm{E_{2_1^+}/E_{break}}$, where $\rm{E_{2_1^+}}$ is the rotor scale determined by the first excited state of the $^{16}$C core and $\rm{E_{break}}$ is set by the energy of its first non-rotational state above the $^{16}$C's ground state. For $^{17}$C, we have: $\rm{E_{2_1^+}/E_{break}} \approx 0.60$. As for $^{10}$Be, this expansion parameter is relatively large and thus may produce slow convergence of the EFT. However, for low-energy processes and states that correspond to an energy $E$ well below the rotor scale the expansion parameter will be $\rm{E/E_{break}}$ which will be markedly smaller than 0.60. 

In particular, the three low-lying bound states of $^{17}$C all correspond to excitation energies of less than 0.5 MeV. They can be reproduced by adjusting the three coupling constants $\rm{C_0}$, $\rm{C_{sd}}$ and $\rm{C_{hf}}$ to its spectrum's properties. Then, we compute Coulomb breakup observables and compare to experimental data. Note that we do not attempt to reproduce 
the $\frac{7}{2}^+$ and $\frac{9}{2}^+$ resonances here, as this would require the explicit inclusion of the 4$^+$ state of $^{16}$C, which goes beyond the scope of this manuscript.  
\begin{table}[t]
\caption{\label{tab:17C_nrjlevels} Table listing the energy levels of the fitted positive parity bound states of $^{17}$C at leading order. The energies and asymptotic normalization coefficients ${\cal C}$ in the $s_{1/2}$ channel are indicated in MeV and $\rm{fm}^{-\frac{1}{2}}$, respectively. The ANCs in the $s$-channel are predictions of our effective theory at leading order. For comparison, experimental data are also displayed \cite{DATTA03,ELE05,BOHL07}. Recent theoretical predictions of ANCs from microscopic $^{17}$C wave functions computed using the resonating group method (RGM) \cite{CHDESC23} are also displayed. The inputs for the EFT calculations are shown in bold, while the predictions are displayed in regular text with their corresponding LO uncertainties.}
\begin{ruledtabular}
\begin{tabular}{cccccccc}
 $\sigma$ (fm) & 1.3 & 1.5 & 2.0 & $\rm{Exp.}$\cite{DATTA03,ELE05,BOHL07} & $\rm{RGM}$ \cite{CHDESC23} \\ \hline \hline 
$\rm{E_{3/2^+}}$ & $\mathbf{-0.7312}$ & $\mathbf{-0.7313}$ & $\mathbf{-0.7312}$  & $-0.73$ & $-0.734$\\
${\cal C}_{s_{1/2} \otimes 2^+}$ & $1.65 \pm 0.87$ & $1.83 \pm 0.96$ & $2.39 \pm 1.25$ & ...  & $1.596$ \\\hline
$\rm{E_{1/2^+}}$ & $\mathbf{-0.5201}$ & $\mathbf{-0.5201}$ & $\mathbf{-0.5200}$ & $-0.52$  &  $-0.517$ \\
${\cal C}_{s_{1/2} \otimes 0^+}$ & $0.77 \pm 0.34$ & $0.81 \pm 0.36$ & $0.92 \pm 0.41$  & ... & $0.959$ \\ \hline
$\rm{E_{5/2^+}}$ & $\mathbf{-0.4033}$ & $\mathbf{-0.4044}$ & $\mathbf{-0.4008}$  & $-0.40$  &   $-0.402$\\
${\cal C}_{s_{1/2} \otimes 2^+}$ & $1.52 \pm 0.59$ & $1.68 \pm 0.65$ & $2.16 \pm 0.84$  & ... &  $0.380$\\ 
\end{tabular}
\end{ruledtabular}
\end{table}
\\

In this context, the ANC in the $s_{1/2}$ channel is a prediction. The energy levels of $^{17}$C calculated at leading order in our theory are shown in Tab.~\ref{tab:17C_nrjlevels}, together with the predicted ANCs. The calculated spectrum reproduces the position of the energy levels of $^{17}$C, regardless of the regulator parameter.  As for $^{11}$Be, the set of defined leading-order operators is sufficient to renormalize binding energies, and indeed the hyperfine operator~(\ref{eq:Vhyp}) is not needed to achieve that renormalization.
Table \ref{tab:17C_nrjlevels} shows that our results exhibit a dependence on the regular parameter $\sigma$ in the predicted $s_{1/2}$ ANCs of the $\frac{3}{2}^+$ ground state, a difference of $\pm 20\%$ between $\sigma$=1.3 and 2.0~fm. This is the expected size of NLO effects. The same applies to the ANC in the $s$-wave channel of the $\frac{1}{2}^+$ and $\frac{5}{2}^+$ excited states. 

As with $^{11}$Be, we compute Coulomb breakup observables involving the $^{17}$C ground state and compare to experimental data. Datta Pramanik and collaborators measured the Coulomb breakup reaction of $^{17}$C on a $^{208}$Pb target: $\rm{^{17}C + Pb\rightarrow}^{16}C(2_1^+)+n+\gamma$ at 495~MeV per nucleon. The measurement of both the decay neutron from excited projectile and $\gamma$-rays emitted from excited fragments after Coulomb breakup allowed them to gate onto the 2$^+$ contribution of the core to this process. 

\begin{figure}[ht]
    \centering
    \begin{minipage}{0.48\linewidth}
        \centering
        \includegraphics[width=\linewidth]
        {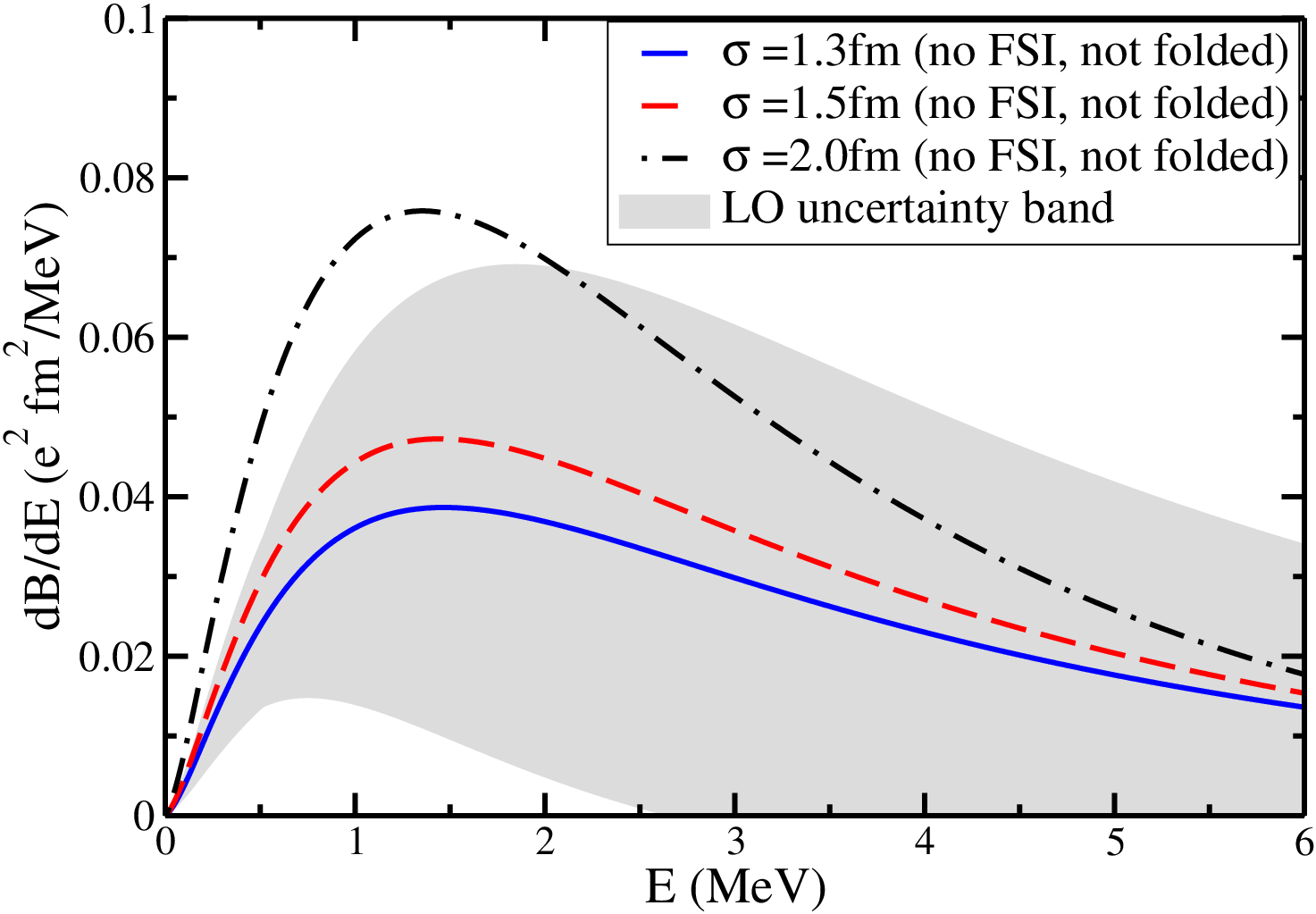}
    \end{minipage}\hfill
    \begin{minipage}{0.48\linewidth}
        \centering
        {\includegraphics[width=\linewidth]
        {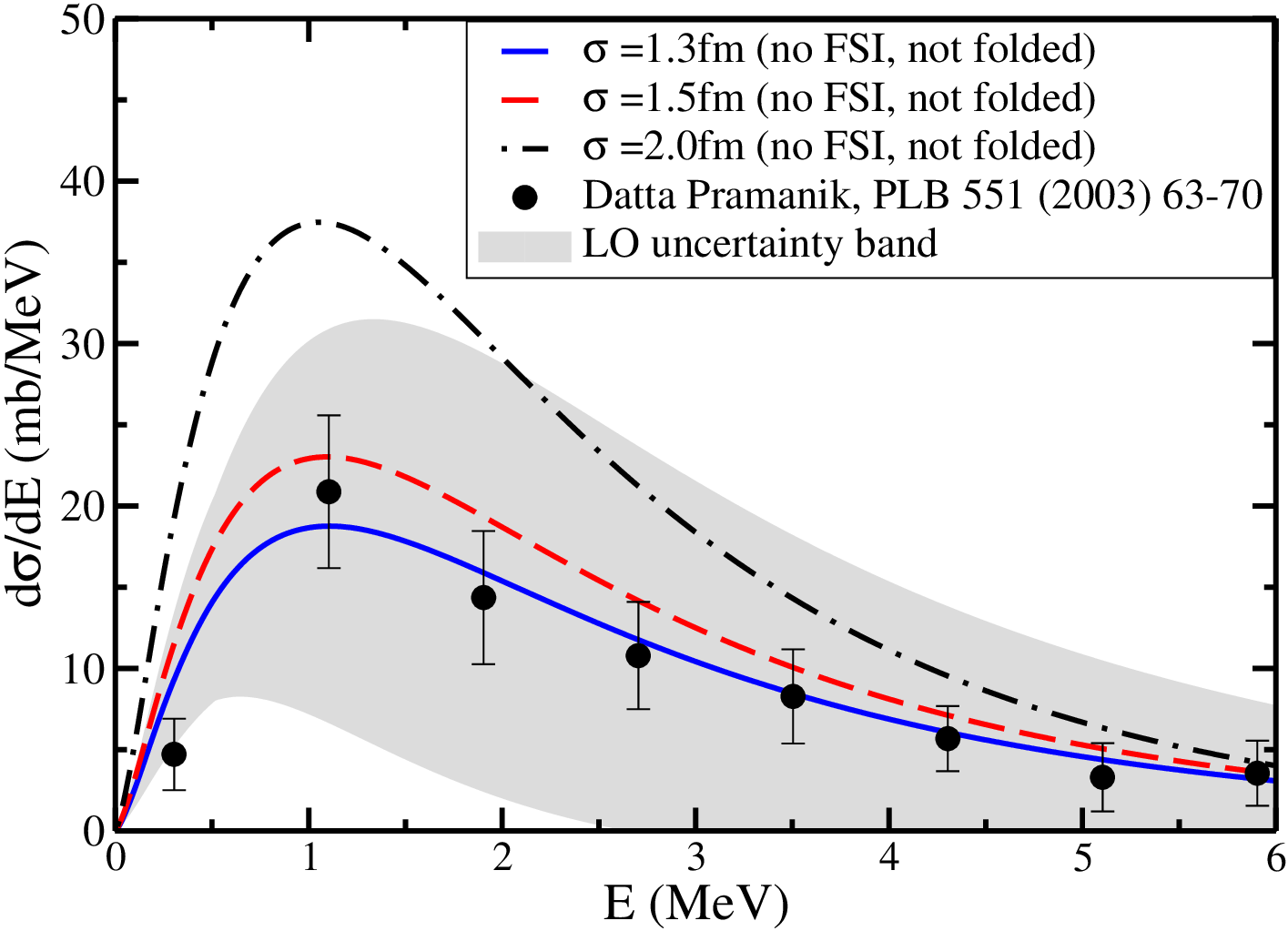}}
    \end{minipage}
    \caption{$\rm{B(E1)}$ distributions (left panel) and Coulomb breakup cross section (right panel) of $^{17}$C: $\rm{^{17}C + Pb\rightarrow}^{16}C(2_1^+)+n+\gamma$ at 495~MeV per nucleon. Our calculations are performed at leading order in our effective theory, without final-state interaction and without folding our results to the experimental resolution. The experimental data, after efficiency and acceptance corrections of all detectors, are also shown for comparison \cite{DATTA03}.}
    \label{fig:dbe1fig17C_LO}
\end{figure}

Once again, using Eq.~(\ref{beTYPBAU}) and the equivalent photon method
 \cite{ALDWIN75,BERBAU88}, we compute the $\rm{B(E1)}$ distributions and the Coulomb breakup cross sections of $^{17}$C. The results are displayed in Fig.~\ref{fig:dbe1fig17C_LO}, considering the $\frac{3}{2}^+$ ground state for the three regulator parameters $\sigma$ in Tab.~\ref{tab:17C_nrjlevels}. These results do not include final-state interactions (FSI) and are not convoluted with the experimental resolution.

In Fig.~\ref{fig:dbe1fig17C_LO}, the regulator-dependence of the s-wave ANCs identified in Tab.~\ref{tab:17C_nrjlevels} is reflected in the breakup observables. Again, this is expected since the Coulomb cross section for loosely bound nuclei scales like the square of the ANC \cite{CAP07}. Thus, while all three regulator parameters are fitted to the same binding energy for the $\frac{3}{2}^+$ ground state, for $\sigma=2.0$~fm, the ANC differs by 50$\%$ compared to the RGM predicted ANC \cite{CHDESC23} and does not reproduce the experimental data at low energies. For $\sigma$=1.3 and 1.5~fm, the ANC differs by a maximum of 15$\%$ from the RGM prediction, and the Coulomb cross sections obtained are in good agreement with the experimental data \cite{DATTA03}. 

As we did for $^{11}$Be above, in Fig.~\ref{fig:dbe1fig17C_LO} we also display the LO uncertainty bands for both the B(E1) distributions and the Coulomb breakup cross sections of $^{17}$C on $^{208}$Pb. Once again, we observe that the majority of our results, obtained with different regulators, lie within the uncertainty bands. The experimental data points are likewise largely covered by the LO uncertainty bands, suggesting that the neglected higher-order effects account for the remaining discrepancy between theory and experiment. We now turn to the consideration of one important NLO effect.
\section{Beyond leading order effects}\label{sec:beyondlo}
In the previous section, we treated $^{11}$Be and $^{17}$C at leading order in our EFT, i.e. at LO in both the halo EFT and rotor expansion parameters $Q$ and $\xi^2$, respectively. This boiled down to solving for the Hamiltonian~(\ref{eq:HLO}).  
In this context, we computed structure and reaction observables, such as ANCs and cross sections and observed that our leading-order calculations had some regulator dependence, the size of which is compatible with NLO effects. Since these observables are peripheral, i.e. sensitive to long-range physics, adding one higher-order finite-range operator should help capturing more long-range physics and reduce regulator dependence.

In this section, we thus propose to study the effects of including one
beyond leading order operator of the form $\rm{C_2(\textbf{p}^2+\textbf{p}'^2)}$ on structure and breakup observables. This is the two-derivative contact operator described in Eq.~(\ref{eq:vNLO}), which corresponds to the derivative of the $s$-wave contact interaction in Eq.~(\ref{eq:vLO}). This operator is of order $O(Q)$ in halo EFT and allows us to now describe $c$-$n$ interactions up to corrections of $O(Q^2)$ in the finite-range Halo EFT expansion. 

Regarding the expansion of the Lagrangian rotor, we keep it as in the last section, i.e. leading order in the rotor expansion parameter $\xi$. Higher-order operators, which are suppressed by $\xi^2$, remain neglected and effects associated with non-rigidity are thus not taken into account in the analysis of this section. 

\subsection{$^{11}$Be}

\begin{figure}[ht] 
    \centering
    \begin{minipage}{0.48\linewidth}
        \centering
        \includegraphics[width=\linewidth]
        {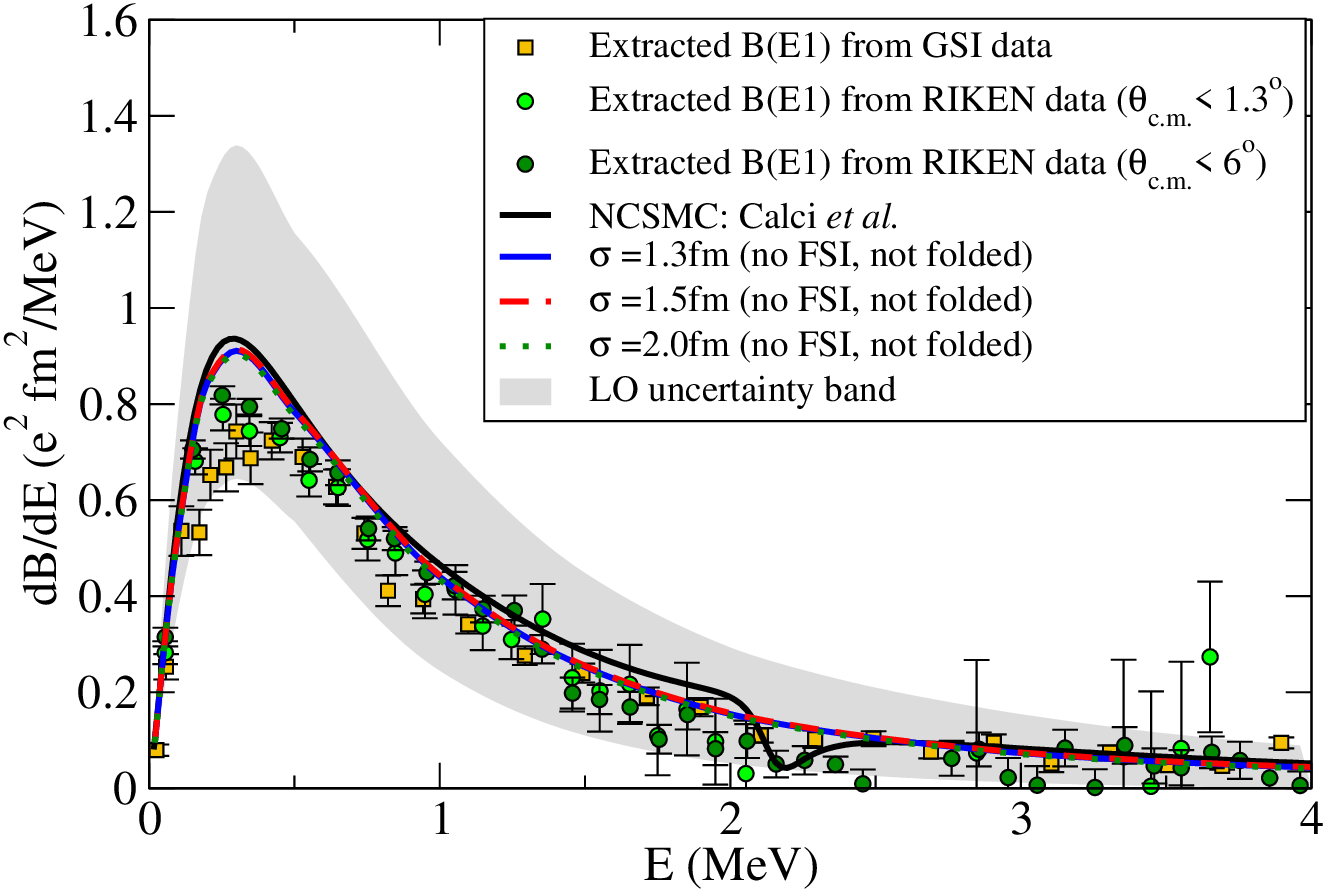}
    \end{minipage}\hfill
    \begin{minipage}{0.48\linewidth}
        \centering
        {\includegraphics[width=\linewidth]
        {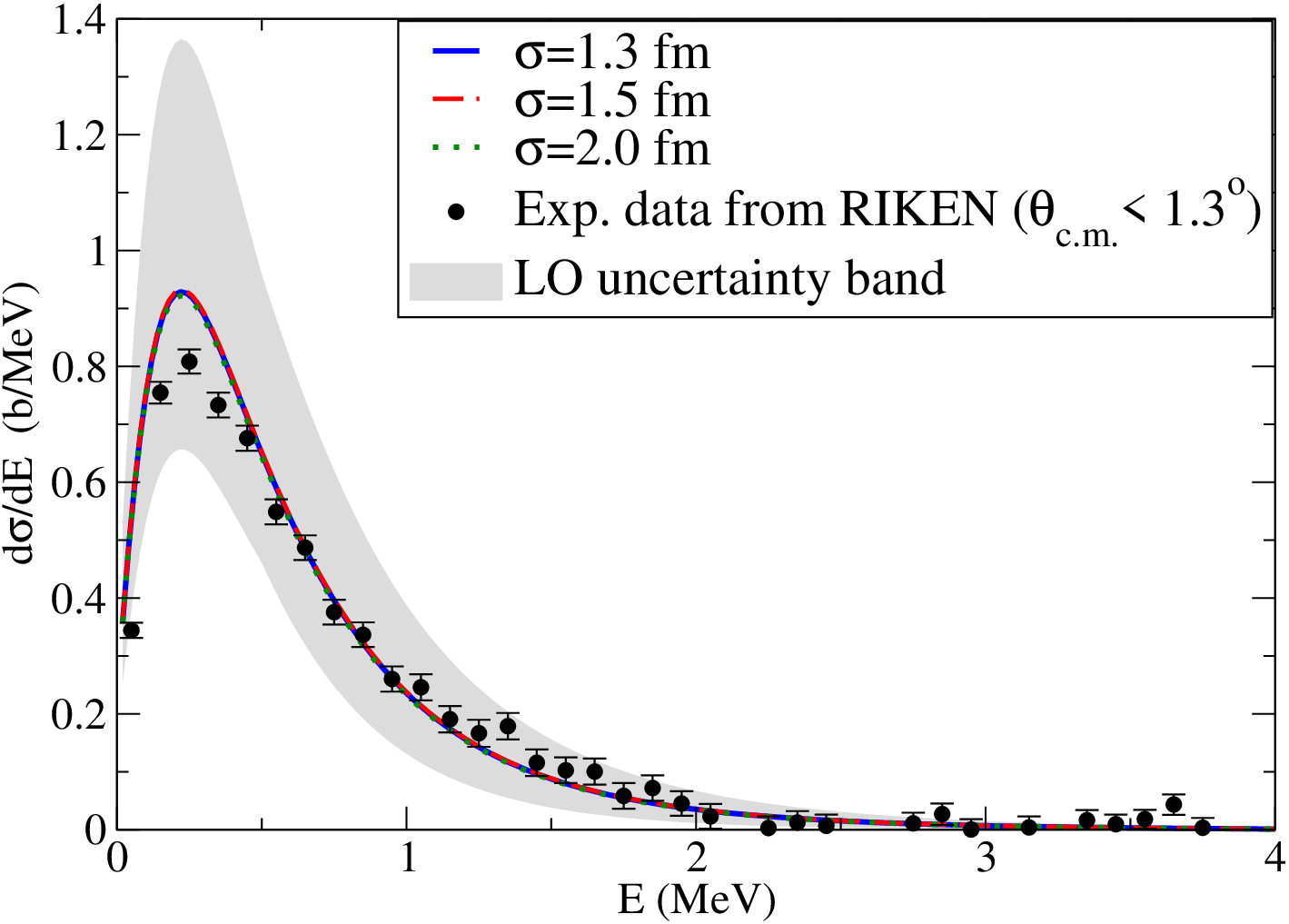}}
    \end{minipage}
    \caption{Beyond leading order calculations of the $\rm{B(E1)}$ distributions (top panel) and the Coulomb breakup cross section (bottom panel) of $^{11}$Be: $\rm{^{11}Be + Pb\rightarrow}^{10}Be+n+Pb$ at 69~MeV per nucleon. Our calculations are performed without final-state interaction in the $p_{3/2}$ partial wave and the breakup cross section is folded with the experimental resolution. Experimental data measured at RIKEN \cite{FUKU04} and GSI \cite{PAL2003} are displayed, as well as \textit{ab initio} NCSMC calculation of Ref.~\cite{CAL16}.}
    \label{fig:dbe1fig11Be_beyondLO}
\end{figure}

For $^{11}$Be, we  renormalize the $s$-wave ANC of the ground state of $^{11}$Be by introducing an operator of the form $\rm{C_2(\textbf{p}^2+\textbf{p}'^2)}$, where $\rm{C_2}$ is a coupling constant that is fitted to reproduce the \textit{ab initio} ANC of its $\frac{1}{2}^+$ ground state \cite{CAL16}. 
By introducing this operator, the spectrum of $^{11}$Be can be recalculated and the properties of the positive parity states obtained are shown in Tab.~\ref{tab:11Be_nrjlevelsbeyondNLO}. As in Sec.~\ref{sec:cases}, we can compute the $\rm{B(E1)}$ distributions as well as the associated Coulomb cross sections, which are displayed in Fig.~\ref{fig:dbe1fig11Be_beyondLO}. It is now clear that, renormalizing the s-wave ANCs leads to regulator-independent $\rm{B(E1)}$ distributions and breakup cross sections and, in addition, a good agreement with the experimental data. For comparison, we also display the LO uncertainty bands (grey bands in Fig.~\ref{fig:dbe1fig11Be_beyondLO}) to compare our partial NLO results to our LO results for the same regulators (see Fig.~\ref{fig:dbe1fig11Be_LO}).

\begin{table}[t]
\caption{\label{tab:11Be_nrjlevelsbeyondNLO} Table summarizing the beyond leading order properties of positive parity states of $^{11}$Be up to the $^{10}$Be($2_1^+$)-$n$ threshold. The energies $\rm{E}$ and widths $\Gamma$ are in MeV and 
the asymptotic normalization coefficients ${\cal C}$ are given in $\rm{fm}^{-\frac{1}{2}}$. The ANC of the $\frac{1}{2}^+$ state of $^{11}$Be has also been computed from \textit{ab initio} No-Core Shell Model with Continuum calculations (NCSMC) and is shown~\cite{CAL16}. Experimental data are also displayed for comparison~\cite{KELL12}. The inputs for the EFT calculations are shown in bold, while the predictions are displayed in regular text with their corresponding LO uncertainties.}
\begin{ruledtabular}
\begin{tabular}{cccccccc}
 $\sigma$ (fm) & 1.3 & 1.5 & 2.0 & $\rm{Exp.}$\cite{KELL12} & $\rm{NCSMC}$ \cite{CAL16} \\ \hline \hline 
$\rm{E_{1/2^+}}$ & $\mathbf{-0.5031}$ & $\mathbf{-0.5031}$ & $\mathbf{-0.5031}$  & $-0.5016$ & $-0.50$\\
${\cal C}_{1/2^+}$ & $\mathbf{0.786}$ & $\mathbf{0.786}$ & $\mathbf{0.786}$ & ... & $0.786$ \\\hline
$\rm{E_{5/2^+}}$ & $\mathbf{1.274}$ & $\mathbf{1.274}$ & $\mathbf{1.274}$ & $1.274$  &  $1.31$ \\
$\Gamma_{5/2^+}$ & $(2.38 \pm 1.33)\times 10^{-3}$ & $(6.68 \pm 3.73)\times 10^{-3}$ & $(3.25 \pm 1.81) \times 10^{-2}$  & $0.100$ & $0.10$ \\ \hline
$\rm{E_{3/2^+}}$ & $\mathbf{2.90}$ & $\mathbf{2.90}$ & $\mathbf{2.90}$  & $2.90$  &   $2.92$\\
$\Gamma_{3/2^+}$ & $0.22 \pm 0.19$ & $0.40 \pm 0.34$ & $1.17 \pm 0.99$  & $0.122$ &  0.06\\ 
\end{tabular}
\end{ruledtabular}
\end{table}
Now, FSI in $p$-waves need to be taken care of. We include some of them, by including FSI in the $p_{1/2}$ partial wave, which hosts a bound state below treshold, while keeping plane waves for the $p_{3/2}$ resonant partial wave. This leads our $\rm{B(E1)}$ and Coulomb cross sections to be in good agreement with experimental data, as well as being cutoff-independent. 

Another noteworthy point is that, compared to our LO calculations, the structure of the $\frac{5}{2}^+$ and $\frac{3}{2}^+$ resonances has not changed much. The predicted widths displayed in Tab.~\ref{tab:11Be_nrjlevelsbeyondNLO} remain strongly regulator-dependent: Correcting the $s$-wave ANC of ${}^{11}$Be states does nothing to remove the short-distance sensitivity seen in this observable. This reinforces the point that renormalizing these widths and incorporating new decay structure effects requires the inclusion of different, higher-order, short-range operators. 
Because the $\frac{5}{2}^+$ and $\frac{3}{2}^+$ resonances of $^{11}$Be are sensitive to $d$-wave interactions, which only enter  at higher order, the prediction for their widths remains consistent with zero, even after the inclusion of the particular NLO effect discussed in this section. 

\subsection{$^{17}$C}
\begin{figure}[hb]
    \centering
    \begin{minipage}{0.48\linewidth} 
        \centering
        \includegraphics[width=\linewidth]
        {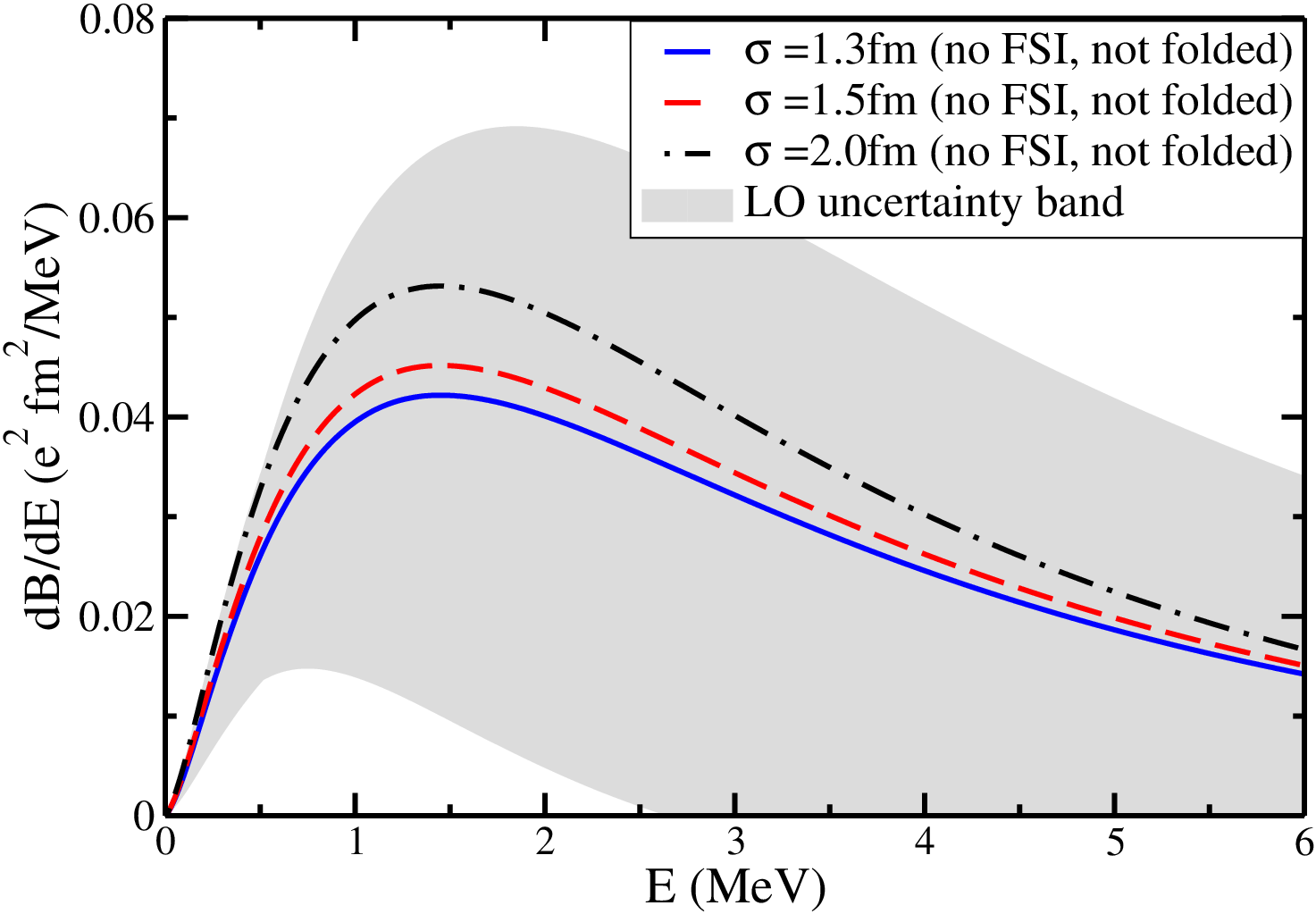}
    \end{minipage}\hfill
    \begin{minipage}{0.48\linewidth}
        \centering
        {\includegraphics[width=\linewidth]
        {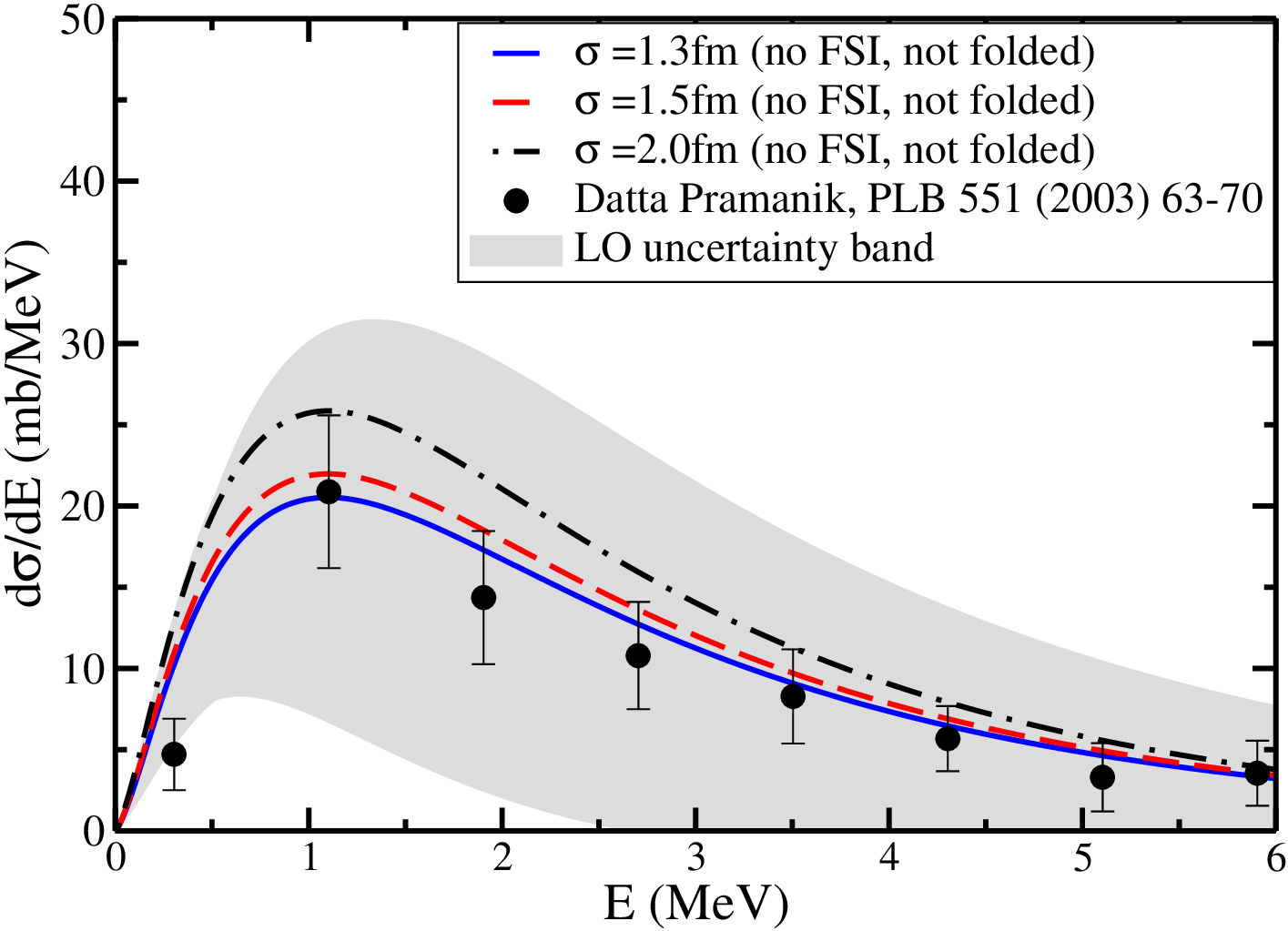}}
    \end{minipage}
    \caption{Beyond leading order calculations of the $\rm{B(E1)}$ distributions (left panel) and Coulomb breakup cross section (right panel) of $^{17}$C: $\rm{^{17}C + Pb\rightarrow}^{16}C(2^+)+n+\gamma$ at 495~MeV per nucleon. Our calculations are performed without final-state interaction and are not folded with the experimental resolution. Experimental data are also shown for comparison \cite{DATTA03}.}
    \label{fig:dbe1fig17C_beyondLO}
\end{figure}

Similarly, the regulator dependence of the $^{17}$C Coulomb breakup cross sections observed at leading order (see Fig.~\ref{fig:dbe1fig17C_LO}) can  be largely eliminated at higher order by introducing relevant higher-order operators. For $^{17}$C, it is also possible to take into account one beyond leading order effect by introducing the $\rm{C_2(\textbf{p}^2+\textbf{p}}^{\prime \, 2})$ operator to renormalize $s$-wave ANCs. The procedure is the same, but we now adjust the low-energy constant $\rm{C_2}$ to stabilize the $s$-wave ANC of the $\frac{1}{2}^+$ halo excited state \cite{CHDESC23}. This ANC was fitted to 0.800~fm$^{-1/2}$, i.e. a value 20$\%$ lower than that predicted by Ref.~\cite{CHDESC23}. Surprisingly, this allows us to renormalize the $s$-wave ANCs of all bound states of $^{17}$C. We now also reproduce the experimental breakup cross section within error bars and significantly reduce regulator-dependence (see Fig.~\ref{fig:dbe1fig17C_beyondLO}). 

Indeed, all $s$-wave ANCs of $^{17}$C's bound states, that are listed in Tab.~\ref{tab:17C_BLO_oper1}, are found to be independent of the regulator. This finding is non-trivial. Indeed, by adjusting only the coupling constant $\rm{C_2}$ associated with the contact two-derivative operator $\rm{C_2(\textbf{p}^2+\textbf{p}}^{\prime \, 2})$ to stabilize the $s$-wave ANC of the $\frac{1}{2}^+$ halo bound state, we can stabilize all $s$-wave ANCs of the other two bound states, $\frac{3}{2}^+$ and $\frac{5}{2}^+$, which was not necessarily expected.

The corresponding beyond leading order calculations of the $\rm{B(E1)}$ distributions and Coulomb cross sections are shown on the left and right panel of Fig.~\ref{fig:dbe1fig17C_beyondLO}, respectively. We also display the LO uncertainty bands (grey bands in Fig.~\ref{fig:dbe1fig17C_beyondLO}) to allow for comparisons between these beyond leading order calculations and our LO results for the same regulators shown in Fig.~\ref{fig:dbe1fig17C_LO}. The residual regulator dependence on the $s$-wave ANCs observed in Tab.~\ref{fig:dbe1fig17C_beyondLO} and on breakup observables (see Fig.~\ref{fig:dbe1fig17C_beyondLO}) is compatible with residual N$^2$LO effects which could be accounted for by a two-derivative hyperfine operator of the form: $\rm{C_2^{hf}(\textbf{p}^2+\textbf{p'}^2)\textbf{I.j}}$. Indeed, for phenomenological reasons (i.e. to reproduce the experimental level ordering), we promoted the hyperfine operator~(\ref{eq:HF}) to LO even though it is not needed to renormalize binding energies. This, in turn, promotes the contribution of the two-derivative hyperfine operator to the Hamiltonian from N$^2$LO to NLO in the EFT.

While the two-derivative $\rm{C_2(\textbf{p}^2+\textbf{p}}^{\prime \, 2})$ operator enables us to stabilize $s$-wave ANCs, it does not allow us to renormalize $d$-wave ANCs, as shown for the $\frac{3}{2}^+$ ground state in Tab.~\ref{fig:dbe1fig17C_beyondLO} (this observation holds for all bound states). These ANCs remain strongly cutoff-dependent. The explicit inclusion of $d$-wave operators to include intrinsic $d$-wave interactions would involve the insertion of operators containing four derivatives and would therefore enter at N$^5$LO in our EFT. This, combined with the fact that the rotor expansion parameter is not small, and so significant subleading corrections are expected, means the widths of these states are not good candidates for an EFT prediction. Nevertheless, this does not invalidate the application of our deformed Halo EFT to ${}^{11}$Be: these $d$-wave properties should be considered to be effectively zero at low orders.

\begin{table}[t]
\caption{\label{tab:17C_BLO_oper1} Evolution of the $s$-wave asymptotic normalization coefficients ${\cal C}$ for all three bound states of $^{17}$C, when including one beyond leading operator (BLO) of the form $\rm{C_2(\textbf{p}^2+\textbf{p'}^2)}$. Its associated LEC $\rm{C_2}$ is adjusted to stabilize the $s$-wave ANC of the $\frac{1}{2}^+$ halo excited state. This value is around 0.80~fm$^{-1/2}$ and allows to both renormalize all $s$-wave ANC and reproduce the experimental breakup cross section within error bars (see right panel of Fig.~\ref{fig:dbe1fig17C_beyondLO}). The inputs for the EFT calculations are shown in bold, while the predictions are displayed in regular text with their corresponding LO uncertainties. Note that the BLO predictions are shown without error bars, since they correspond to partial NLO calculations.
}
\begin{ruledtabular}
\begin{tabular}{cccccccc}
 $\sigma$ (fm) & 1.3 & 1.5 & 2.0  \\ \hline \hline 
$\frac{1}{2}^+ - {\cal C}_{s1/2 \otimes 0^+}$ - \textbf{LO} & $0.77 \pm 0.34$ & $0.81 \pm 0.36$ & $0.92 \pm 0.41$
\\ 
$\frac{1}{2}^+ - {\cal C}_{s1/2 \otimes 0^+}$  \textbf{BLO}& $\mathbf{0.80}$ & $\mathbf{0.80}$ & $\mathbf{0.80}$
\\ \hline \hline 
$\frac{3}{2}^+ - {\cal C}_{s_{1/2} \otimes 2^+}$ - \textbf{LO} & $1.65 \pm 0.87$ & $1.83 \pm 0.96$ & $2.39 \pm 1.25$  
\\ 
$\frac{3}{2}^+ - {\cal C}_{s_{1/2} \otimes 2^+}$ \textbf{BLO} & $1.74$ & $1.78$ & $1.87$
\\ \hline \hline
$\frac{5}{2}^+ - {\cal C}_{s_{1/2} \otimes 2^+}$ - \textbf{LO} & $1.52 \pm 0.59$ & $1.68 \pm 0.65$ & $2.16 \pm 0.84$ 
\\ 
$\frac{5}{2}^+ - {\cal C}_{s_{1/2} \otimes 2^+}$  \textbf{BLO} & $1.60$ & $1.64$ & $1.71$  
\\ \hline \hline
$\frac{3}{2}^+ - {\cal C}_{d_{3/2} \otimes 0^+}$ - \textbf{LO}& $(1.2 \pm 0.64)\times 10^{-2}$ & $(1.7 \pm 0.90)\times 10^{-2}$ & $(3.5 \pm 1.82)\times 10^{-2}$  
\\ 
$\frac{3}{2}^+ - {\cal C}_{d_{3/2} \otimes 0^+}$ \textbf{BLO} & $1.3 \times 10^{-2}$ & $1.7 \times 10^{-2}$ & $ 3.5 \times 10^{-2}$  
\end{tabular}
\end{ruledtabular}
\end{table}

These beyond leading order calculations give us an indication of the possible order of magnitude of the corrections to the ANC that would come into play at next-to-leading order and higher orders in the EFT. This also informs us about the order-by-order  renormalizability of the theory. Indeed, we have shown that the remaining regulator dependence seen in the leading-order $\rm{B(E1)}$ distributions is contained in the renormalization of the $s$-wave ANCs (in particular that of the $\frac{3}{2}^+$ ground state), i.e. in long-range physics.
\section{Conclusion}
In this paper, we have constructed a short-range effective theory for deformed $s$-wave one-neutron halo nuclei, i.e., halos in which the core nucleus exhibits a low-lying rotational band  with a $0^+$ ground state and a first $2_1^+$ excited state. Exploiting the separation of scales associated with both halo nuclei and rotational band, we included degrees of freedom associated with both low-energy excitations. At leading order the resulting EFT reproduces the particle plus rotor model  \cite{BOMOT69} with a contact potential governing the neutron-core interaction.

First, we reviewed the key ingredients of Halo EFT, and then we defined how it can be extended to deformed halos by incorporating rotational core degrees of freedom, along with a new associated expansion parameter. We then introduced the power counting and identified the relevant leading-order operators in the Hamiltonian that respect the symmetries of the system.

Second, we solved for the low-lying positive-parity EFT energy levels of core$+$neutron systems up to the core$(2_1^+)$ treshold for several different regulator parameters. We showed that their binding energies can be renormalized using the defined leading-order set of operators. We then computed Coulomb breakup observables and compared them with experimental data. Having performed these calculations for $^{11}$Be and $^{17}$C, we showed that their spectra can be accurately reproduced at leading order. 

For $^{11}$Be, we found that the decay widths to $d$-wave core-neutron states have sizable cutoff dependence, but the $s$-wave asymptotic normalization coefficients exhibit regulator dependence of a size consistent with next-to-leading-order effects. Similar results were obtained for $^{17}$C, with its $d$-wave ANCs found to be strongly regulator dependent. We also noted that leading-order $\rm{B(E1)}$ strength Coulomb breakup cross sections for both $^{11}$Be and $^{17}$C are in fair agreement with experimental data. Like $s$-wave ANCs, they exhibit some cutoff dependence, but its size is compatible with next-to-leading order effects. 

For $^{11}$Be and $^{17}$C, we also studied one beyond leading order effect by including one finite-range two-derivative operator in our EFT. We showed that, by stabilizing the single $s$-wave ANC associated with their $\frac{1}{2}^+$ halo state using one finite-range two-derivative operator, E1 breakup observables can be made almost entirely cutoff-independent while being in good agreement with experimental data. In addition, for $^{17}$C, including this one operator allows us to renormalize of all its $s$-wave ANCs, i.e. the $s$-waves ANCs associated with its $\frac{3}{2}^+$ ground and $\frac{5}{2}^+$ second excited state.

Several avenues are open to build upon these first steps towards the establishment of our effective theory. First, the construction of full next-to-leading order and its application to the cases of $^{11}$Be and $^{17}$C is needed to firmly establish that this EFT is a successful and systematic expansion for these systems. Construction of the full NLO Hamiltonian is in progress. 
The EFT could also be applied to $^{19}$C, which is another halo nucleus whose core exhibits rotational low-lying $0^+$ and $2_1^+$ states. Furthermore, we plan to extend this “deformed halo EFT" to p-wave (negative-parity) states, which will make possible an EFT treatment of $p$-wave driven,  halos that are built on a deformed core, such as $^{31}$Ne. 
Also note that, in its current form, our effective theory could also be employed to describe one-proton halos with a deformed core, such as ${}^{22}$Al~\cite{CBETAL24}. Finally, following our approach, an EFT could potentially be developed in the future to describe two-nucleons halo systems with a deformed core.
\section*{Acknowledgments}
We thank U.~Datta Pramanik for sending us the Coulomb breakup experimental data for $^{17}$C of Ref.~\cite{DATTA03}, after efficiency and acceptance correction of all detectors. This work was supported by the U.S.~Department of Energy under contract No.~DE-FG02-93ER40756.
\bibliography{bibliography}
\end{document}